\documentclass{my_iopjournal}

\usepackage{amsmath}
\usepackage{amssymb}
\usepackage{multicol}
\usepackage{ragged2e}
\usepackage{dcolumn}% Align table columns on decimal point
\usepackage{bm}% bold math
\usepackage[utf8]{inputenc}
\usepackage[T1]{fontenc}
\usepackage{mathptmx}
\usepackage{float}
\usepackage{etoolbox}
\usepackage{siunitx}
\usepackage{booktabs}
\usepackage{xfrac}
\usepackage[colorlinks=true, allcolors=blue]{hyperref}
\usepackage{threeparttable}
\usepackage{titlesec}
\usepackage[backend=bibtex, style=numeric-comp, sorting=none, maxnames=99]{biblatex}
\DeclareSIUnit{\dBm}{dBm}

\titlelabel{\thetitle.\enskip}

\begin{document}

\title{Effects of the thin-film thickness on superconducting NbTi microwave resonators for on-chip cryogenic thermometry}

\author{
André Chatel\textsuperscript{\textcolor{blue}{1},\textcolor{blue}{2},\textcolor{blue}{*}}\orcid{0000-0003-3016-0104}, 
Roberto Russo\textsuperscript{\textcolor{blue}{1},\textcolor{blue}{2}}\orcid{0000-0002-6489-888X}, 
Luca Mazzone\textsuperscript{\textcolor{blue}{1}}, 
Quentin Boinay\textsuperscript{\textcolor{blue}{1}}, 
Reza Farsi\textsuperscript{\textcolor{blue}{1}}\orcid{0000-0002-8257-7254}, 
Jürgen Brugger\textsuperscript{\textcolor{blue}{1}}\orcid{0000-0002-7710-5930}, 
Giovanni Boero\textsuperscript{\textcolor{blue}{1},\textcolor{blue}{2}}\orcid{0000-0003-4913-3114}, 
Hernan Furci\textsuperscript{\textcolor{blue}{1}}\orcid{0000-0003-0595-4180}
}

\affil{\textsuperscript{1}Microsystems Laboratory, \'Ecole Polytechnique Fédérale de Lausanne (EPFL), 1015 Lausanne, Switzerland}

\affil{\textsuperscript{2}Center for Quantum Science and Engineering, \'Ecole Polytechnique Fédérale de Lausanne (EPFL), 1015 Lausanne, Switzerland}

\affil{\textsuperscript{*}Author to whom any correspondence should be addressed.}

\email{\textcolor{blue}{andre.chatel@epfl.ch}}

\keywords{Superconducting thin-films, Microwave resonators, Kinetic inductance, Cryogenic thermometry, Quantum sensing}

\begin{abstract}
\vspace{3mm} 
\justifying
\noindent
Superconducting microwave resonators have recently gained a primary importance in the development of cryogenic applications, such as circuit quantum electrodynamics, electron spin resonance spectroscopy and particles detection for high-energy physics and astrophysics. In this work, we investigate the influence of the film thickness on the temperature response of microfabricated Nb\textsubscript{50}Ti\textsubscript{50} superconducting resonators. S-shaped split ring resonators (S-SRRs), $\text{\SI{20}{\nano\metre}}$ to $\text{\SI{150}{\nano\metre}}$ thick, are designed to be electromagnetically coupled with standard Cu coplanar waveguides (CPWs) and their microwave properties are characterized at temperatures below $\text{\SI{10}{\kelvin}}$. The combined contributions of the kinetic inductance $L_{K}(T)$ increase and the decreasing loaded quality factor $Q_{L}$, for thinner films, induce an optimum condition on the temperature sensitivity and resolution of the resonators. A noise equivalent temperature ($NET$) as low as $\text{\SI{0.5}{\mu\kelvin/\hertz^{1/2}}}$, at $\text{\SI{1}{\hertz}}$, is reported for $\text{\SI{100}{\nano\metre}}$ thick resonators at $\text{\SI{4.2}{\kelvin}}$. We also asses the possibility of implementing a multiplexed frequency readout, allowing for the simultaneous temperature tracking of several sensors along a single CPW. Such results demonstrate the possibility to perform a distributed cryogenic temperature monitoring, with a sub-$\text{\SI{}{\mu\kelvin}}$ resolution. Thus, the application of superconducting S-SRRs, eventually benefiting from an even higher $L_K(T)$, for a further miniaturization, as well as a back-end integration directly on-chip, can be envisioned for the accurate monitoring of localized temperature of devices operating in cryogenic conditions.
\end{abstract}

\begin{multicols*}{2}
\justifying

\section{Introduction}
In the last decades, microwave systems operating at temperatures well below $\text{\SI{90}{\kelvin}}$ have experienced a growing interest from the scientific community. In this framework, superconducting thin-films have paved the way for novel techniques and devices relying on low temperature quantum phenomena. Among different technologies, superconducting resonators have proven to be key elements for the design of microwave kinetic inductance detectors (MKIDs) \cite{Audley:1993, Day:2003, Maloney:2010, Leduc:2010, Austermann:2018, Dibert:2022}, circuit quantum electrodynamics (QED) architectures \cite{Wallraff:2004, Frunzio:2005, Goppl:2008, Scarlino:2019} and qubits \cite{Kjaergaard:2020, Koch:2007, Schreier:2008, Barends:2014}, ultra-low noise quantum parametric amplifiers \cite{Vissers:2016, Aumentado:2020, Frasca:2024} and detectors for electron spin resonance (ESR) spectroscopy \cite{Bienfait:2016, Ranjan:2020, Artzi:2021, Akhmetzyanov:2023, Russo:2025}. Dealing with all the aforementioned applications, the accurate monitoring of the on-chip temperature is crucial for ensuring the correct functioning of the device. The distributed temperature monitoring, in large scale cryogenic apparatuses, is a well-established technology, with commercially-available cryogenic thermometers, as CERNOX\textsuperscript{\textregistered} \cite{Courts:2003, Courts:2014} or RuO\textsubscript{2} \cite{Myers:2021} thermistors, generally showing wide $\text{\SI{1}{}-\SI{300}{\kelvin}}$ operational ranges, large temperature sensitivities and temperature resolutions around $\text{\SI{100}{\mu\kelvin}}$, below $\text{\SI{10}{\kelvin}}$ \cite{Holmes:1992, Yeager:2001, Ekin:2010}. However, the size of such devices is, most of the times, comparable to the dimensions of cryogenic chips and the power dissipation, as well as the need for individual electrical routings, introduce important design limitations when allocating several sensors to perform localized on-chip temperature monitoring of complex quantum architectures. In such a context, superconducting materials have been recently investigated as alternatives to traditional technologies, with DC CMOS-compatible devices \cite{Noah:2024} and microwave resonators \cite{Wheeler:2020} having reported temperature resolutions, respectively, around $\text{\SI{3}{\milli\kelvin/\hertz^{1/2}}}$ and $\text{\SI{75}{\mu\kelvin/\hertz^{1/2}}}$, for $T < \text{\SI{1}{\kelvin}}$. More specifically, the superconducting properties in the microwave range can be exploited to realize resonators able to correlate changes in the cryogenic temperature to frequency shifts \cite{Olivadese:2019, Wheeler:2020, Yu:2022}, with the potential of performing a temperature sensing in parallel with the operation of the chip. High quality factors, larger than $\text{10\textsuperscript{4}}$, multiplexed readouts through a single microwave coplanar line, the possibility of a direct back-end integration on-chip and the ease of performing an accurate and multiple temperature monitoring make superconducting microwave resonators an attractive alternative to more conventional cryo-thermistors \cite{Wheeler:2020, Olivadese:2019}.

In this work, we report on the dependence of the temperature sensitivity and resolution of superconducting microresonators on the superconducting film thickness. The aim of this study is to provide guidelines for the optimization of the performances of cryogenic thermometers based on the variation of the thin-film kinetic inductance. Our findings demonstrate that thinner superconducting films simultaneously entail (a), as an advantage, an increase in the temperature sensitivity of the resonators, through the increase of $L_K(T)$, and (b), as a drawback, an increase in the minimum detectable temperature variations, as a result of the reduction of the quality factor. Thus, when thinning down the devices, a trade-off between these two figures-of-merit (FOMs) has to be carefully evaluated in the design phase, in order to determine an optimal film thickness. In this perspective, the best resolution is achieved for $\text{\SI{100}{\nano\metre}}$ thick resonators, for which an ultimate noise equivalent temperature ($NET$) as good as $\text{\SI{0.5}{\mu\kelvin/\hertz^{1/2}}}$, at $\text{\SI{1}{\hertz}}$ and $\text{\SI{4.2}{\kelvin}}$, is recorded, out-performing the existing available technologies by at least a factor $\text{\SI{100}{}}$ (see the \textcolor{blue}{supplementary data, S1}). Finally, the possibility to implement a simultaneous and frequency-multiplexed temperature monitoring, with a sub-$\text{\SI{}{\mu\kelvin}}$ resolution, is successfully demonstrated by applying frequency modulation techniques to two separate resonators, individually excited by the same microwave coplanar line.

\section{Methods}

The kinetic inductance $L_K(T)$ of superconducting materials is an electronic-transport parameter originating from the low inertia of Cooper pairs, whose density $n_S(T)$ significantly decreases for temperatures approaching the superconducting-to-normal critical transition \cite{Gorter:1934, Watanabe:1994, Yoshida:1995, Hein:2001, Tinkham:2004, Frunzio:2005}. In particular, considering a thin-film superconducting resonator, with length $l$, width $w$ and normal state surface resistance $R_{\square}(T_C)$, it is possible to model the temperature influence on its resonance frequency, close to the transition temperature $T_C$ and in the low frequency regime $\hbar\omega \ll k_BT$, as \cite{Tinkham:2004, Annunziata:2010, Nulens:2023}
\begin{subequations}
\label{EQN_01}
\begin{equation}
f_{res}(T) = \frac{1}{2\pi\sqrt{(L_G+L_K(T)) C_G}}
\label{EQN_01a}
\end{equation}
\begin{equation}
L_K(T) = \frac{\hbar R_{\square}(T_C)}{\pi \Delta(T)} \frac{1}{\tanh\Bigl(\frac{\Delta(T)}{2k_BT}\Bigr)} \biggl(\frac{l}{w}\biggr)
\label{EQN_01b} 
\end{equation} 
\end{subequations} 
where $\Delta(T) \simeq 1.74 \Delta(0)\sqrt{1-\sfrac{T}{T_C}}$ is the superconducting energy gap and $ \Delta(0) = 1.76 k_BT_C$, while $L_G$ and $C_G$ stand, respectively, for the geometric inductance and capacitance of the resonator. Such a temperature dependence $f_{res}(L_K(T))$ is the key physical phenomenon exploited to realize the the superconducting thin-film microwave resonators presented in this work.

\begin{figure*}[!ht]
\includegraphics[width=\textwidth]{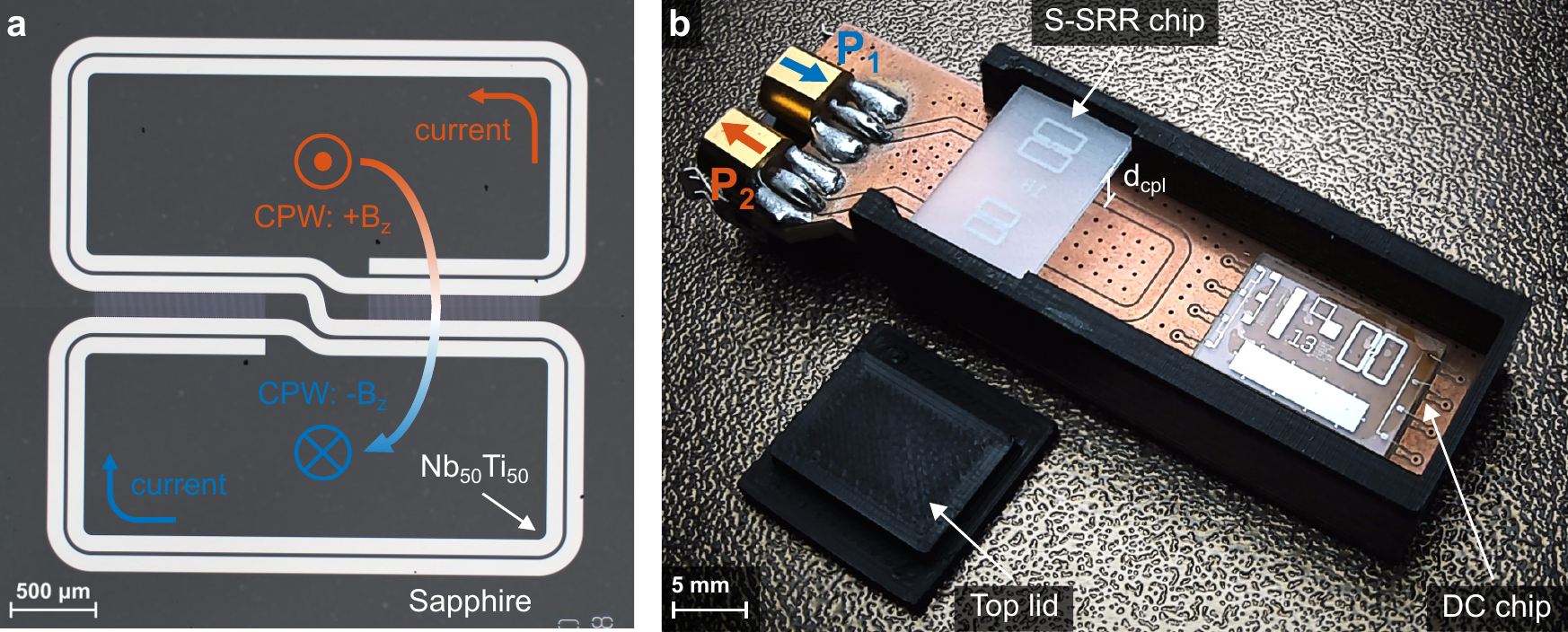}
\caption{\justifying Description of the S-SRR chip and excitation PCB. a) Optical microscope picture of the $\text{\SI{1.1}{\giga\hertz}}$ resonator under test. The coloured arrow represents the direction of the antisymmetric B-field of a GCPW underneath. b) PCB exploited for the characterization of the resonators in cryogenic environments. The separation distance between the chip and the Cu GCPW is set by the black PLA support; a plastic lid (bottom-left) and Kapton-tape are exploited to keep the suspended chip in position. The presence of two resonators, showing a similar geometry but a significant difference in overall sizes (and, thus, in resonance frequency), has to be attributed to our initial intention to perform a simultaneous frequency-multiplexed readout in different regions of the microwave spectrum. A chip containing 4-wires structures, for the Nb\textsubscript{50}Ti\textsubscript{50} films cryogenic DC characterization, is wire-bonded at the bottom-right of the PCB.}
\label{PANEL_01}
\makeatletter
\let\save@currentlabel\@currentlabel
\edef\@currentlabel{\save@currentlabel(a)}\label{PANEL_01:a}
\edef\@currentlabel{\save@currentlabel(b)}\label{PANEL_01:b}
\makeatother
\end{figure*}

\subsection{Selection of the materials}
In order to study the influence of the film thickness on the temperature response of superconducting resonators, several Nb\textsubscript{50}Ti\textsubscript{50} films, with the thickness ranging from $\text{\SI{20}{\nano\metre}}$ to $\text{\SI{150}{\nano\metre}}$, are DC sputtered, at room temperature, on C-plane sapphire wafers. The choice of Nb\textsubscript{50}Ti\textsubscript{50} lies in its transition parameters, showing critical DC field values as high as $\text{\SI{15}{\tesla}}$ \cite{Bychkov:1981, Bottura:2000, Ghigo:2023} and microwave ones larger than $\text{\SI{350}{\milli\tesla}}$ \cite{Russo:2023}, which can reduce the cross-sensitivity of the kinetic inductance-based sensors to magnetic field fluctuations. Moreover, a bulk critical temperature up to $\text{\SI{10}{\kelvin}}$ \cite{Bychkov:1981, Benvenuti:1991, Charifoulline:2006, Zhang:2019} makes this alloy compatible with our cryogenic setup and suitable for temperature ranges commonly achieved by cryogenic applications based on liquid He (LHe) cooling. Sapphire is exploited\hfill as\hfill substrate\hfill material\hfill because\hfill of\hfill its\hfill exceptional\hfill mi-
\begin{table}[H]
\centering
\begin{threeparttable}
\caption{\justifying Nb\textsubscript{50}Ti\textsubscript{50} critical surface resistance, $\text{\SI{0}{\kelvin}}$ kinetic inductance, critical temperature, geometric inductance and capacitance of the resonators, for different thin-film thicknesses.}
\label{TABLE_01}
\renewcommand{\arraystretch}{1.25}
\begin{tabular*}{\columnwidth}{@{\hspace{10pt}\extracolsep{\fill}} cccccc @{\hspace{10pt}}}
\toprule
$t$ & $R_\square\text{($T_C$)}$\tnote{\textcolor{blue}{a}} & $L_K\text{(0)}$ & $T_C$\tnote{\textcolor{blue}{b}} & $L_G$\tnote{\textcolor{blue}{b}} & $C_G$\tnote{\textcolor{blue}{b}} \\
$\text{(nm)}$ & $\text{ ($\Omega$/}\square\text{)}$ & $\text{(pH/}\square\text{)}$ & $\text{(\SI{}{\kelvin})}$ & $\text{(\SI{}{\nano\henry})}$ & $\text{(\SI{}{\femto\farad})}$\\
\midrule
24 & 51 & 11 & 6.2 & 48 & 440\\
57 & 17 & 3.2 & 7.7 & 35 & 560\\
105 & 8.7 & 1.5 & 8.1 & 29 & 660\\
147 & 5.7 & 1.0 & 8.1 & 27 & 710\\
\bottomrule
\end{tabular*}
\begin{tablenotes}
\item[a] \footnotesize{Extracted from a 4-wires V/I measurement.}
\item[b] \footnotesize{Fitting parameters from the $L_K(T)$-based model presented in \autoref{EQN_01}.}
\end{tablenotes}
\end{threeparttable}
\end{table}
\noindent crowave \cite{Buckley:1994, Krupka:1999, Pogue:2012} and thermal \cite{Dobrovinskaya:2009, Brown:2010} properties in ultra-cold environments, as, for instance, a high dielectric constant $\epsilon_r$ around $10$, an extremely low loss factor $\epsilon''$, down to $10^{-9}$ and a thermal conductivity $\kappa_T$ even larger than $\text{\SI{1000}{\watt/m\cdot\kelvin}}$, below $\text{\SI{80}{\kelvin}}$. A lower bound for the surface kinetic inductance of the Nb\textsubscript{50}Ti\textsubscript{50} films is estimated through the $L_{K}(T)$ at $T = \text{\SI{0}{\kelvin}}$ \cite{Tinkham:2004, Yu:2021}, with the value of the different normal state surface resistances measured just above the critical transition by means of a 4-wires patterned structure. In \autoref{TABLE_01} we report the estimated surface kinetic inductances. A maximum of $\SI{11}{}$ pH/$\square$ is obtained for the $\text{\SI{24} {\nano\metre}}$ thick film. In the framework of optimising the temperature-sensing performances, the use of nitride-based superconducting thin-films, such as NbN and NbTiN \cite{Yu:2021, Parker:2022, Frasca:2023} could, in principle, lead to larger $L_{K}(0)$ and, eventually, to better sensitivity and resolution when used for cryogenic thermometry.

\begin{table*}
\centering
\begin{threeparttable}
\caption{\justifying Resonance frequency $f_{res}$, loaded $Q_L$, coupling $Q_c$ and internal $Q_i$ quality factors, phase mismatch $\phi$ and goodness of fit $R^2$ coefficient, estimated on the $\text{\SI{147}{\nano\metre}}$ thick S-SRR, at $\text{\SI{4.2}{\kelvin}}$, for different coupling distances. A decrease in the internal Q-factor, for a stronger coupling, is observed, probably attributed to the small, but not negligible, microwave resistance of the Cu GCPW inductive coupling section of about $\text{\SI{3}{\milli\metre}}$.}
\label{ADD_INFO_TABLE_01}
\renewcommand{\arraystretch}{1.25}
\begin{tabular*}{\textwidth}{@{\hspace{10pt}\extracolsep{\fill}} ccccccc @{\hspace{10pt}}}
\toprule
$d_{cpl}$ (mm) & $f_{res}$ (GHz) & $Q_{L}$ ($\times$10\textsuperscript{4}) & $Q_{c}$ ($\times$10\textsuperscript{6}) & $Q_{i}$ ($\times$10\textsuperscript{4}) & $\phi$ (\textdegree) & $R^2$ (\%)\\
\midrule
1.5 & 1.1533 & 0.68 & 0.01 & 2.16 & 1.02 & 99.999\\
2.5 & 1.1531 & 1.64 & 0.05 & 2.43 & 2.50 & 99.996\\
3.0 & 1.1548 & 2.08 & 0.11 & 2.58 & 1.93 & 99.998\\
4.0 & 1.1534 & 2.65 & 1.06 & 2.72 & 2.34 & 99.999\\
5.0 & 1.1537 & 2.75 & 3.17 & 2.77 & 2.73 & 99.999\\
\bottomrule
\end{tabular*}
\end{threeparttable}
\end{table*}

\subsection{Design of the sensing resonators}
Among different planar geometries, S-shaped split ring resonators (S-SRRs) have been reported as ideal candidates for providing an optimal electromagnetic coupling with microwave coplanar waveguides (CPWs) \cite{Horestani:2014, Herrojo:2016, Horestani:2019}, due to the possibility of exciting two counter-twisted inductive loops with antisymmetric magnetic field configurations [see \autoref{PANEL_01:a}]. In such a perspective, Nb\textsubscript{50}Ti\textsubscript{50} S-SRRs are patterned on $\text{\SI{650}{\mu\metre}}$ thick sapphire wafers, by recurring to a combination of direct laser writing (DLW) photolithography and SF\textsubscript{6}/CHF\textsubscript{3}-based plasma etching. More details about the micro-fabrication of the thin-film resonators are reported in the \textcolor{blue}{supplementary data, S2}, where an assessment of the patterning quality is also provided. The S-SRR geometry consists of 2-turns top and bottom loops ($\text{\SI{3.20}{\milli\metre}}$ $\times$ $\text{\SI{1.52}{\milli\metre}}$). The inductive stripe shows length and width values, respectively, equal to $\text{\SI{37}{\milli\metre}}$ and $\text{\SI{100}{\mu\metre}}$. An inter-line gap of $\text{\SI{20}{\mu\metre}}$ separates each turn, while two interdigitated capacitors (IDCs), made up of $\text{\SI{340}{}}$ fingers (length: $\text{\SI{148}{\mu\metre}}$, width: $\text{\SI{4}{\mu\metre}}$, gap: $\text{\SI{2}{\mu\metre}}$), are located between the two loops. Such additional capacitors are intended to maintain the compactness of the device, while lowering down its resonance frequency and limiting, therefore, the degradation of the quality factor induced by the $\propto f^2$ increase of the superconducting surface resistance \cite{Tinkham:2004, Hein:2001, Mansour:2002}.

COMSOL Multiphysics\textsuperscript{\textregistered} finite element method (FEM) and Quasi Universal Circuit (QUC) simulations are performed to optimize the S-SRR geometry for coupling inductively to a standard Cu grounded CPW (GCPW) and resonating approximatively around $\text{\SI{1.1}{\giga\hertz}}$ (see the \textcolor{blue}{supplementary data, S3}). The chips are coupled face-down, by geometric proximity, to a $\text{\SI{50}{\ohm}}$ matched GCPW by means of a 3D-printed polylactide (PLA) support, allowing to precisely place the S-SRRs with the two inductive loops centered on top of the transmission line gaps, as shown in \autoref{PANEL_01:b}. In such a configuration, the critical coupling parameter turns out to be the distance $d_{cpl}$ between the chip and the printed circuit board (PCB).

\subsection{Determination of the coupling distance}

\begin{figure*}[t!]
\includegraphics[width=\textwidth]{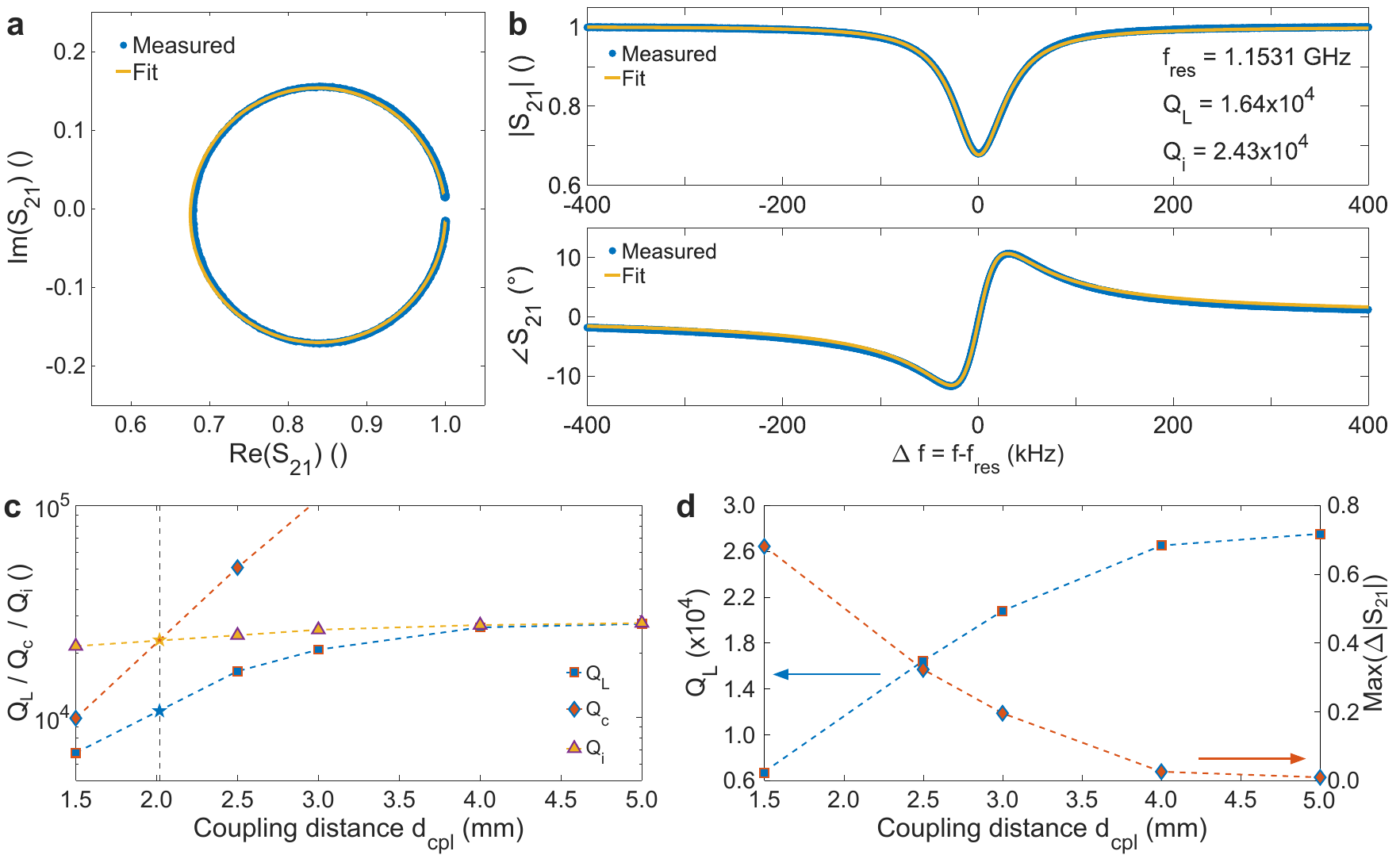}
\caption{\justifying Experimental search for an optimal resonator-to-GCPW coupling distance. a) Renormalized real and imaginary parts of the $S_{21}$ parameter, for the $d_{cpl} = \text{\SI{2.5}{\milli\metre}}$ case. The complex data (blue dots) are fitted to a notch-type circular model (yellow line). b) Magnitude and phase of the $S_{21}$ $S_{21}$ parameter, for the $d_{cpl} = \text{\SI{2.5}{\milli\metre}}$ case. c) Evolution of the loaded $Q_L$ (dashed blue line), coupling $Q_c$ (dashed orange line) and internal $Q_i$ (dashed yellow line) quality factors with respect to the coupling distance. The single pentagram points represent the critical coupling condition (\textit{i.e.} $Q_i = Q_c \longrightarrow Q_L = Q_i/2$ ). d) Evolution of the loaded Q-factor (dashed blue line) and the resonance peak depth (dashed orange line), for an increasing coupling distance.}
\label{PANEL_02}
\makeatletter
\let\save@currentlabel\@currentlabel
\edef\@currentlabel{\save@currentlabel(a)}\label{PANEL_02:a}
\edef\@currentlabel{\save@currentlabel(b)}\label{PANEL_02:b}
\edef\@currentlabel{\save@currentlabel(c)}\label{PANEL_02:c}
\edef\@currentlabel{\save@currentlabel(d)}\label{PANEL_02:d}
\edef\@currentlabel{\save@currentlabel(a-b)}\label{PANEL_02:a-b}
\edef\@currentlabel{\save@currentlabel(c-d)}\label{PANEL_02:c-d}
\makeatother
\end{figure*}

In order to experimentally identify an optimal coupling condition for our resonators, a $\text{\SI{147}{\nano\metre}}$ thick device is preliminary tested at a fixed $\text{\SI{4.2}{\kelvin}}$ temperature, inside a liquid He dewar. The same resonator chip is successively characterized at different coupling distances $d_{cpl}$ from the same GCPW PCB, by changing the 3D-printed spacer. The system is excited in a 2-port scheme directly through a vector network analyzer (VNA), with a $\text{-\SI{18}{\dBm}}$ input power, performing a frequency sweep to locate the resonance condition on the transmission $S_{21}$ parameter. The extraction of the quality factors, for each $d_{cpl}$ value, is performed by fitting the complex-plane data to a notch-type resonator model \cite{Gao:2008, Khalil:2012, Probst:2015, Nigro:2024, De_Palma:2024}
\begin{equation}
\label{EQN_03}
S_{21}(f) = a e^{i(\alpha-2\pi\tau f)}\biggl( 1-\frac{(Q_L/|Q_c|)e^{i\phi}}{1+2iQ_L(f/f_{res}-1)} \biggr)
\end{equation}
where the term $a e^{i(\alpha-2\pi\tau f)}$ accounts for the environmental effects introduced by the excitation microwave cabling (\textit{i.e.} $a$ stands for the amplitude attenuation, while $\alpha$ for the phase shift and $\tau$ for the line delay) and $\phi$ takes into consideration impedance mismatches at the input/output ports. Moreover, $f_{res}$ represents the resonance frequency of the S-SRR, while $Q_L$ and $Q_c$ denote, respectively, the loaded and coupling/external quality factors. These latter are linked, together with the $Q_i$ internal/unloaded Q-factor of the resonator, by the constitutive relation $Q_L^{-1} = Q_i^{-1} + Q_c^{-1}$. The estimation of the resulting parameters is reported in \autoref{ADD_INFO_TABLE_01}, together wit the goodness of fit $R^2$ coefficient, typically larger than $99.99 \%$. The coincidence between the measured $S_{21}$ data and the complex-plane circular fit, upon the environmental pre-factor re-normalization, is shown in \autoref{PANEL_02:a-b}, for the specific case of $d_{cpl} = \text{\SI{2.5}{\milli\metre}}$. Moreover, \autoref{PANEL_02:c} reports on the evolution of the Q-factors with respect to the coupling distance. The saturation of $Q_L$ to the unloaded condition occurs for a $Q_{L}(\SI{5}{\milli\metre}) = \text{\SI{2.75}{}$\times$10\textsuperscript{4}}$, while the critical coupling distance is estimated around $\text{\SI{2}{\milli\metre}}$. Considering also the evolution peak depth [see \autoref{PANEL_02:d}], a slightly under-coupling distance of $\text{\SI{2.5}{\milli\metre}}$ is chosen for further variable-temperature characterizations. This choice is motivated by the possibility to clearly isolate and distinguish resonance peaks, with loaded Q-factor values larger than half the saturation condition, still preserving an energy transfer, at resonance, larger than $50\%$ (\textit{i.e.} $1-|S_{21}(f_{res})|^2 = 1-0.68^2 \simeq 54\%$).

\section{Results and discussion}
In the following sections, we report on the main experiments carried out to determine the temperature sensitivity and resolution of the the S-SRRs, for different film thicknesses (\textit{i.e.} $\text{\SI{147}{\nano\metre}}$, $\text{\SI{105}{\nano\metre}}$, $\text{\SI{57}{\nano\metre}}$ and $\text{\SI{24}{\nano\metre}}$). Additionally, we also present the data related to the simultaneous temperature monitoring of two distinct resonators (\textit{i.e.} $\text{\SI{147}{\nano\metre}}$ and $\text{\SI{105}{\nano\metre}}$) in liquid He, demonstrating the possibility to perform an accurate frequency-multiplexed temperature readout. In the end, we investigate the origin of the very low-frequency signal fluctuations that seem to limit the estimation of the $NET$ below $\text{\SI{1}{\hertz}}$. This last experiment is further validated by assessing the S-SRR readout against the temperature monitoring performed by means of a commercially-available RuO\textsubscript{2} thermistor, providing a direct comparison between the performances of our resonators and a state-of-the-art cryogenic thermometer.

\subsection{Estimation of the temperature sensitivity}
\begin{figure}[H]
\includegraphics[width=\linewidth]{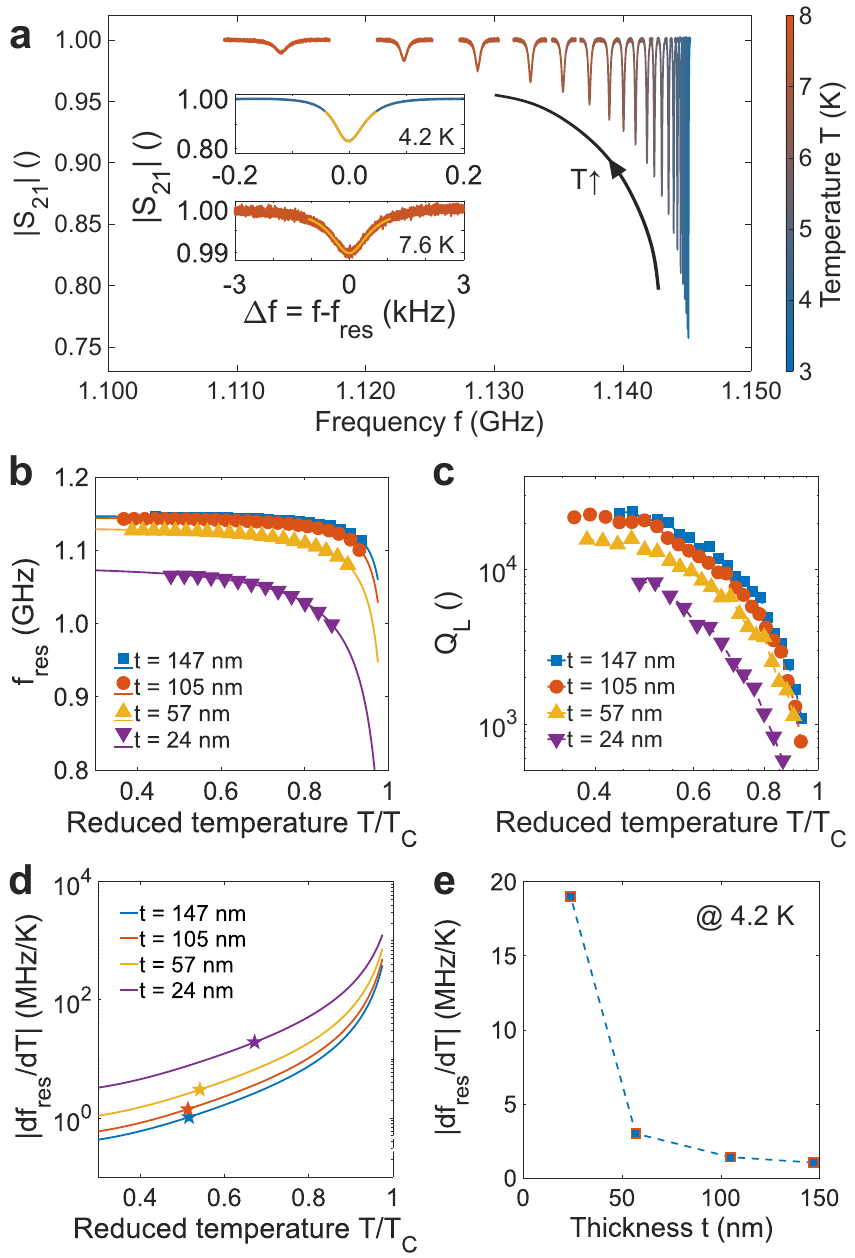}
\caption{\justifying Temperature response of the S-SRRs for different film thicknesses. a) Temperature-induced resonance frequency shifts for the $\text{\SI{147}{\nano\metre}}$ thick sample. The two insets show the $|S_{21}|$ resonances at $\text{\SI{4.2}{\kelvin}}$ ($f_{res} = \text{\SI{1.1442}{\giga\hertz}}$, $Q_{L} = \text{\SI{1.61}{}$\times$10\textsuperscript{4}}$) and $\text{\SI{7.6}{\kelvin}}$  ($f_{res} = \text{\SI{1.1133}{\giga\hertz}}$, $Q_{L} = \text{\SI{1.08}{}$\times$10\textsuperscript{3}}$): the yellow curves stand for the notch-type complex plane circular fit. b) Resonance frequency shifts for the different film thicknesses: the solid line represents the $f_{res}(T)$ model, fitted to the data points. c) Temperature evolution of the loaded Q-factor. d) Temperature evolution of the S-SRRs sensitivity, estimated as the derivative of the $f_{res}(T)$ fitting model. The single points refer to the value at $\text{\SI{4.2}{\kelvin}}$. e) Temperature sensitivity of the different S-SRRs, evaluated at $\text{\SI{4.2}{\kelvin}}$.}
\label{PANEL_03}
\makeatletter
\let\save@currentlabel\@currentlabel
\edef\@currentlabel{\save@currentlabel(a)}\label{PANEL_03:a}
\edef\@currentlabel{\save@currentlabel(b)}\label{PANEL_03:b}
\edef\@currentlabel{\save@currentlabel(c)}\label{PANEL_03:c}
\edef\@currentlabel{\save@currentlabel(d)}\label{PANEL_03:d}
\edef\@currentlabel{\save@currentlabel(e)}\label{PANEL_03:e}
\edef\@currentlabel{\save@currentlabel(d-e)}\label{PANEL_03:d-e}
\makeatother
\end{figure}

In order to determine the temperature response of the resonators for different thicknesses, the S-SRRs are cooled down, in zero-field conditions, to temperatures as low as $\text{\SI{3}{\kelvin}}$, by means of the variable temperature insert (VTI) of a dry superconducting cryomagnet. A mitigation of the devices cross-sensitivity to remanent $B$-field fluctuations is implemented by encapsulating the whole system, comprising of the sensing S-SRR and excitation PCB, in a cylindrical magnetic shield made of CRYOPHY\textsuperscript{\textregistered}, a Ni-Fe soft magnetic alloy from \textit{Magnetic Shields Ltd}, typically exploited to shield magnetic fields in cryogenic environments ($\mu_r$ around $ \text{\SI{7}{}$\times$10\textsuperscript{4}}$ at $\text{\SI{4}{\kelvin}}$). A 2-port VNA, directly connected in transmission to the PCB, is exploited to excite each device with a power of $\text{\SI{-18}{\dBm}}$, allowing to record the S-parameters of the system. The temperature monitoring and control is achieved through the use of two CERNOX\textsuperscript{\textregistered} thermometers and two heaters (see the \textcolor{blue}{supplementary data, S4}). In \autoref{PANEL_03:a} we report the typical $S_{21}$ spectral response below $T_C$. The effect of the microwave lines is eliminated by acquiring frequency scans\hfill at\hfill a\hfill temperature\hfill slightly\hfill above\hfill the\hfill superconducting
\noindent transition and subtracting such backgrounds, in the complex domain, to the corresponding resonance signals \cite{Hafner:2014, Rausch:2018}. The values of the resonance frequency $f_{res}$ and loaded $Q_L$, coupling $Q_c$ and internal $Q_i$ quality factors are, then, extracted by fitting the complex  S\textsubscript{21} transmission parameter to the notch-type resonator model from \autoref{EQN_03}. The temperature response of the different S-SRRs is compared in \autoref{PANEL_03:b}, with the data-points being fitted to the $f_{res}(T)$ model from \autoref{EQN_01}. The resonance frequency of each device decreases as the temperature approaches the critical transition, consistently with the increase of the kinetic inductance term $L_K(T)$ \cite{Watanabe:1994, Yoshida:1995, Frunzio:2005}. Moreover, an overall shift towards lower frequencies for $f_{res}(0)$ is observed, which can probably be attributed to the increase of the geometric inductance $L_G$, induced by the smaller cross-section of the thinner films. As described by \autoref{TABLE_01}, the increase in the $L_G$ term is well reflected\hfill by\hfill the\hfill evolution\hfill of\hfill such\hfill a\hfill fitting\hfill parameter, as\hfill well
\begin{figure}[H]
\includegraphics[width=\linewidth]{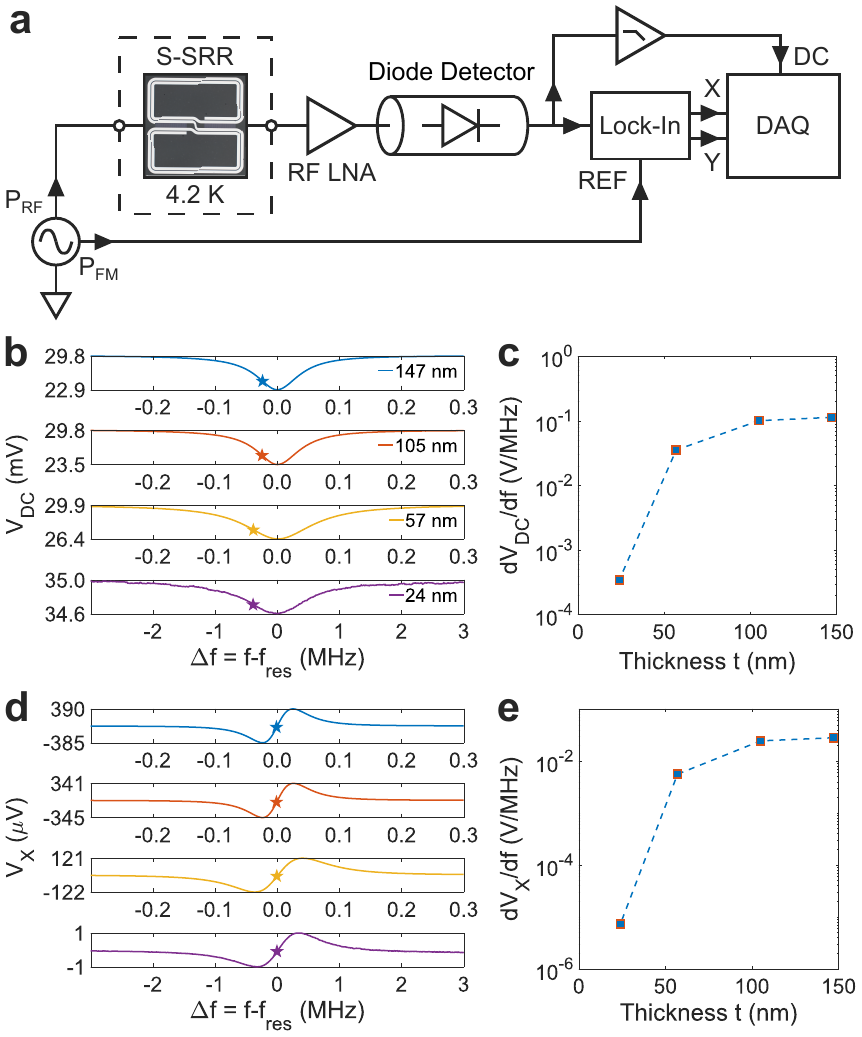}
\caption{\justifying Frequency response of the S-SRRs for different film thicknesses. a) Experimental setup used to characterize the devices in liquid He at $\text{\SI{4.2}{\kelvin}}$ b) DC component of the resonance signal: the single points represent the maximum slope condition. c) Maximum slope of the DC component versus film thickness. d) X component of the lock-in output signal: the single points represent the maximum slope condition, in this case located at the resonance frequency. The color legend is the same as the one reported for the DC signal graph. e) Maximum slope of the X component versus film thickness.}
\label{PANEL_04}
\makeatletter
\let\save@currentlabel\@currentlabel
\edef\@currentlabel{\save@currentlabel(a)}\label{PANEL_04:a}
\edef\@currentlabel{\save@currentlabel(b)}\label{PANEL_04:b}
\edef\@currentlabel{\save@currentlabel(c)}\label{PANEL_04:c}
\edef\@currentlabel{\save@currentlabel(d)}\label{PANEL_04:d}
\edef\@currentlabel{\save@currentlabel(e)}\label{PANEL_04:e}
\edef\@currentlabel{\save@currentlabel(b-c)}\label{PANEL_04:b-c}
\edef\@currentlabel{\save@currentlabel(d-e)}\label{PANEL_04:d-e}
\makeatother
\end{figure}
\noindent as the critical temperature $T_C$ follows the saturation trend reported in literature for thicker films \cite{Takeda:1999, Inomata:2009, Bretz-Sullivan:2022}. Additionally, the loaded quality factor $Q_L$, even though showing values as high as $\text{2.5$\times$10\textsuperscript{4}}$, is found to experience a decrease with respect to the reduction of the S-SRRs thickness, most probably due to the larger microwave resistance of the thinner films [see \autoref{PANEL_03:c}]. The temperature sensitivity of each resonator is also estimated, by means of the $f_{res}(T)$ analytical model. The result is reported in \autoref{PANEL_03:d-e} together with the evidence of the higher overall sensitivity for the devices characterized by larger values of kinetic inductance. For instance, an almost 20-fold larger sensitivity value of $\text{\SI{19}{\mega\hertz/\kelvin}}$, is achieved, with respect to the $\text{\SI{147}{\nano\metre}}$ thick S-SRR, for the $\text{\SI{24}{\nano\metre}}$ thick one, at $\text{\SI{4.2}{\kelvin}}$.

\subsection{Estimation of the temperature resolution}
\label{SECTION_A}
The temperature resolution of such superconducting devices is estimated at $\text{\SI{4.2}{\kelvin}}$ in a LHe dewar, in order to eliminate all sources of environmental noise, possibly related to remanent magnetic field fluctuations and gaseous-He temperature instabilities in our VTI system \cite{Russo:2025}. Additionally, a more complex experimental setup, based on a frequency modulation (FM) lock-in detection scheme, is implemented [see \autoref{PANEL_04:a}]. A microwave signal generator is exploited to excite the S-SRRs with a $\text{\SI{-20}{\dBm}}$ power, frequency-modulated at $\text{\SI{11}{\kilo\hertz}}$ with a deviation of $\pm\text{\SI{5}{\kilo\hertz}}$. The choice of these three values is justified in the \textcolor{blue}{supplementary data, S5}. The microwave transmitted signal is amplified and converted into a DC value proportional to its amplitude with a microwave diode detector. The different components of the signal are then split, with the DC one further amplified, and the $\text{\SI{11}{\kilo\hertz}}$ AC one detected by means of a lock-in amplifier, with a $\text{\SI{2.5}{\hertz}}$ equivalent noise bandwidth (\textit{i.e.} $\text{\SI{100}{\milli\second}}$ time constant). Finally, both DC, in-phase (X) and quadrature (Y) components are simultaneously acquired by a data acquisition (DAQ) board with a $\text{\SI{100}{\milli\second}}$ integration time (more details  are presented in  the \textcolor{blue}{supplementary data, S5}). Such a characterization setup is preferred to more conventional schemes, relying on VNAs \cite{Gao:2008b, Wheeler:2020, Yu:2022} and homodyne readout systems \cite{Gao:2007, Kumar:2008}, to investigate the possible application of signal modulation techniques for reducing the influence of the instrumentation noise at lower frequencies. In \autoref{PANEL_04:b-c} we report the frequency response of the DC component of the resonance signal. The dependance of the loaded quality factor $Q_L$ on the film thickness, already observed in the previous variable-temperature experiments, affects the slope of the resonance curves. In particular, when thinning down the Nb\textsubscript{50}Ti\textsubscript{50} film from $\text{\SI{147}{\nano\metre}}$ to $\text{\SI{24}{\nano\metre}}$, we observe a reduction of $Q_L$ by a factor of 20 and a reduction of the slope by a factor of 300 (from $1.1\times10^{-1}\text{ \SI{}{\volt/\mega\hertz}}$ to $3.5\times10^{-4}\text{ \SI{}{\volt/\mega\hertz}}$). Such a strong reduction of $Q_L$ also impacts the response of the resonators in terms of the signal X component [see \autoref{PANEL_04:d-e}], where limiting values of $2.8\times10^{-2}\text{ \SI{}{\volt/\mega\hertz}}$ and $7.3\times10^{-6}\text{ \SI{}{\volt/\mega\hertz}}$ define the maximum slope conditions for the thickest and thinnest devices respectively.

\begin{figure}[H]
\includegraphics[width=\linewidth]{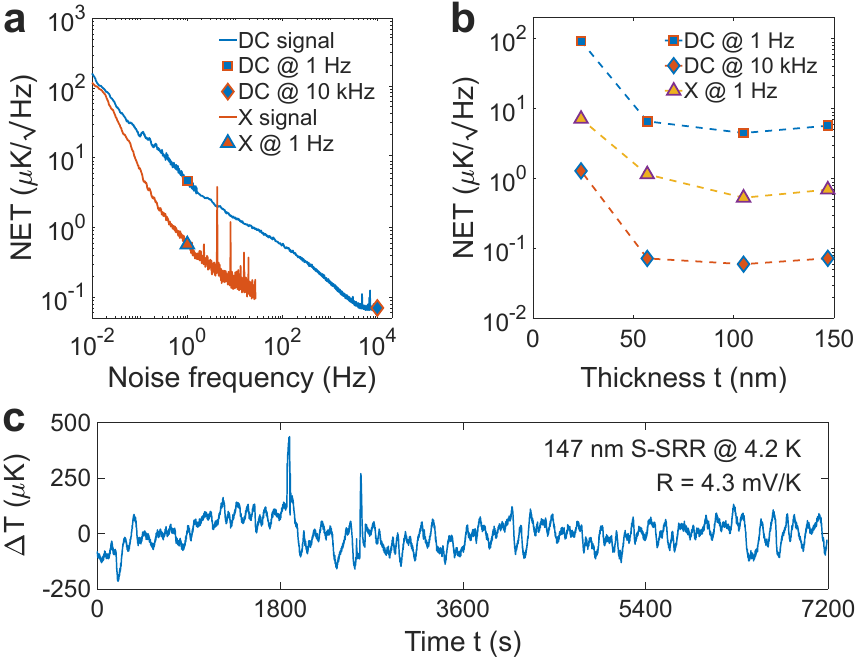}
\caption{\justifying Noise characterization of the different S-SRRs at $\text{\SI{4.2}{\kelvin}}$. a) Frequency spectra of the $NET$ for the $\text{\SI{147}{\nano\metre}}$ thick resonator. b) Noise equivalent temperatures of the different S-SRRs, evaluated in DC, both at $\text{\SI{10}{\kilo\hertz}}$ (blue-square points) and $\text{\SI{1}{\hertz}}$ (orange-diamond points), and in AC, at $\text{\SI{1}{\hertz}}$ (yellow-triangle points). c) Typical $\text{\SI{2}{\hour}}$ time evolution of an S-SRR temperature monitoring through the X signal (in this case, related to the $\text{\SI{147}{\nano\metre}}$ thick resonator).}
\label{PANEL_05}
\makeatletter
\let\save@currentlabel\@currentlabel
\edef\@currentlabel{\save@currentlabel(a)}\label{PANEL_05:a}
\edef\@currentlabel{\save@currentlabel(b)}\label{PANEL_05:b}
\edef\@currentlabel{\save@currentlabel(b)}\label{PANEL_05:c}
\makeatother
\end{figure}

The temperature resolution, or noise equivalent temperature ($NET$), expressed in $\text{\SI{}{\kelvin/\hertz^{1/2}}}$, is defined as
\begin{subequations}
\label{EQN_02}
\begin{equation}
NET(f_{noise}) = \frac{NSD(f_{noise})}{R}
\label{EQN_02a}
\end{equation}
\begin{equation}
R = \dfrac{dV}{dT} = \dfrac{dV}{df} \cdot \dfrac{df}{dT}
\label{EQN_02b}
\end{equation}
\end{subequations}
where $NSD(f_{noise})$ is the noise spectral density (in $\text{\SI{}{\volt/\hertz^{1/2}}}$), while $R$ is the temperature responsivity of the device (in $\text{\SI{}{\volt/\kelvin}}$) as the product between the slope of the sensing signal (\textit{i.e} either DC or X) and the temperature sensitivity of the resonance frequency. The low-frequency NSDs, below $\text{\SI{40}{\hertz}}$, are extracted by applying the fast Fourier transform (FFT) on a $\text{\SI{2}{\hour}}$ time monitoring of both the DC and X signals, with $\text{\SI{5}{\milli\second}}$ lock-in and DAQ integration times. \autoref{PANEL_05:a} shows the $NET$ frequency spectrum for the $\text{\SI{147}{\nano\metre}}$ thick S-SRR, estimated by means of \autoref{EQN_02}. The $NET$ associated to the DC signal clearly shows a $1/f$ behavior below $\text{\SI{10}{\kilo\hertz}}$, which might be associated to the electronic noise coming from the setup instrumentation. In particular, a temperature resolution of about $\text{\SI{6}{\mu\kelvin/\hertz^{1/2}}}$ is achieved for such a signal at $\text{\SI{1}{\hertz}}$. The FM transfer of the sensing signal above the $1/f$ corner frequency, minimizes the influence of the instrumentation noise and allows to improve the temperature resolution of the device almost by a factor 10, with a $NET$ around $\text{\SI{0.7}{\mu\kelvin/\hertz^{1/2}}}$ for the X signal at $\text{\SI{1}{\hertz}}$. The higher temperature sensitivity of\hfill thinner\hfill devices\hfill is\hfill counterbalanced\hfill by\hfill the\hfill degradation\hfill of 
\noindent the quality factor (\textit{i.e.} sensing curve slope $dV/df$). As previously discussed, these two effects contribute in originating an optimum condition for the best temperature resolution. Considering the X signal evaluated at $\text{\SI{1}{\hertz}}$, a minimum $NET$ of about $\text{\SI{0.5}{\mu\kelvin/\hertz^{1/2}}}$, among the investigated film thicknesses, is obtained for the $\text{\SI{105}{\nano\metre}}$ thick resonator [see \autoref{PANEL_05:b}]. Surprisingly, the temperature resolution below $\text{\SI{1}{\hertz}}$ seems to be limited by low-frequency fluctuations in the order of $\text{\SI{100}{\mu\kelvin}}$, as shown in \autoref{PANEL_05:c}. The presence of atomic two-levels systems (TLSs), at the interface between glassy substrates and superconducting materials \cite{Muller:2019} could, in principle affect the noise estimation of the resonators. Nevertheless, such phenomena should become significant at much lower temperatures than the ones considered in our work (\textit{i.e.} for $T < T_C/8$), as reported by other literature studies \cite{Kumar:2008, Gao:2008b, Wheeler:2020}.

\subsection{Investigation of the low-frequency fluctuations}
\begin{figure*}[t!]
\includegraphics[width=\textwidth]{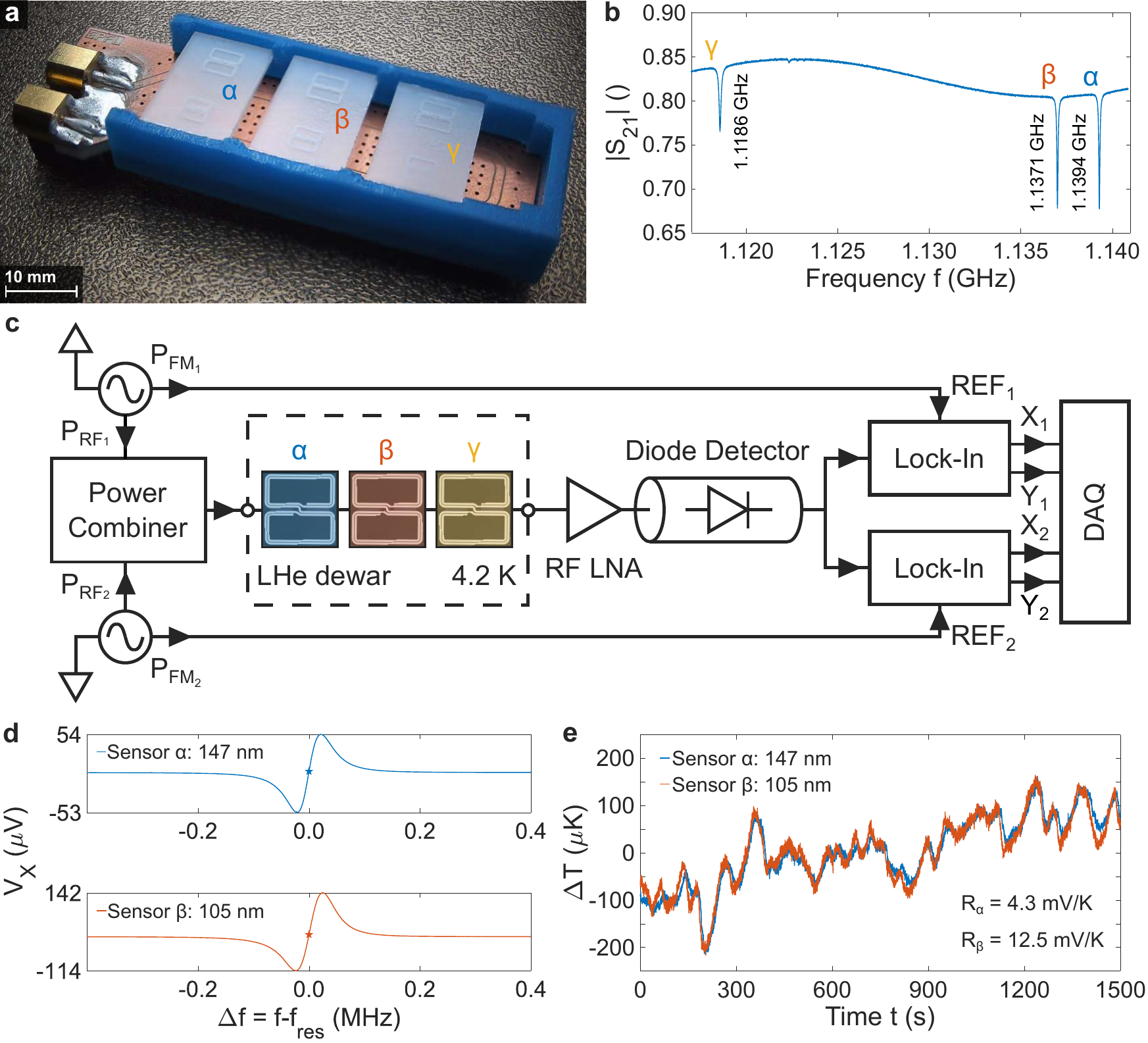}
\caption{\justifying Two resonators simultaneous temperature monitoring at $\text{\SI{4.2}{\kelvin}}$ in a LHe dewar. a) PCB exploited to excite multiple resonators, by means of a single Cu GCPW. b) $S_{21}$ transmission parameter frequency scan, in liquid He at $\text{\SI{4.2}{\kelvin}}$. From left to right, it is possible to clearly distinguish the presence of the three different resonance peaks, respectively related to the $\text{\SI{57}{\nano\metre}}$,  $\text{\SI{105}{\nano\metre}}$ and  $\text{\SI{147}{\nano\metre}}$ thick S-SRRs. c) Experimental setup used to simultaneously excite two-out-of-three S-SRRs, by means of two microwave signals frequency-modulated at two different values. d) Simultaneous frequency scan of the X signals for the $\text{\SI{147}{\nano\metre}}$ and $\text{\SI{105}{\nano\metre}}$ thick devices (Sensor $\alpha$: $f_{res} = \text{\SI{1.1394}{\giga\hertz}}$ and $Q_L = \text{\SI{1.67}{}$\times$10\textsuperscript{4}}$; Sensor $\beta$: $f_{res} = \text{\SI{1.1371}{\giga\hertz}}$ and $Q_L = \text{\SI{1.61}{}$\times$10\textsuperscript{4}}$). The single points represent the zero-crossing resonance frequency, showing the maximum slope values. e) $\text{\SI{25}{\minute}}$ time-lapse evolution of the two X signals, simultaneously exciting the two S-SRRs at their corresponding resonance frequency.}
\label{PANEL_06}
\makeatletter
\let\save@currentlabel\@currentlabel
\edef\@currentlabel{\save@currentlabel(a)}\label{PANEL_06:a}
\edef\@currentlabel{\save@currentlabel(b)}\label{PANEL_06:b}
\edef\@currentlabel{\save@currentlabel(c)}\label{PANEL_06:c}
\edef\@currentlabel{\save@currentlabel(d)}\label{PANEL_06:d}
\edef\@currentlabel{\save@currentlabel(e)}\label{PANEL_06:e}
\makeatother
\end{figure*}

The origin of such fluctuations is, therefore, investigated by means of an experiment where two resonators are simultaneously measured (\textit{i.e.} the $\text{\SI{147}{\nano\metre}}$ and the $\text{\SI{105}{\nano\metre}}$ thick ones). \autoref{PANEL_06:a} shows the Cu PCB used to excite three different S-SRRs in parallel, resonating around $\text{\SI{1.1}{\giga\hertz}}$. The 3D-printed PLA support is re-designed in order to host three chips (\textit{i.e.} the $\text{\SI{147}{\nano\metre}}$,  $\text{\SI{105}{\nano\metre}}$ and  $\text{\SI{57}{\nano\metre}}$ thin-films), $\text{\SI{2.5}{\milli\metre}}$ far from the GCPW and approximatively spaced by $\text{\SI{11}{\milli\metre}}$. The exact value of each resonance frequency, at $\text{\SI{4.2}{\kelvin}}$, is preliminarily determined by performing a frequency scan with a VNA, delivering an excitation power of $\text{\SI{-20}{\dBm}}$. Each of the peaks is well separated from the others by several $\text{\SI{}{\mega\hertz}}$, allowing to unambiguously identify a different excitation frequency for each of them [see \autoref{PANEL_06:b}].

Once having located the different resonances, the two S-SRRs are individually excited by the same GCPW at their corresponding resonance frequencies (i.e. $\text{\SI{1.1394}{\giga\hertz}}$ and $\text{\SI{1.1371}{\giga\hertz}}$, respectively). The schematic reporting this more complex setup is illustrated in \autoref{PANEL_06:c}. Two different signal generators are exploited to provide distinct $\text{-\SI{20}{\dBm}}$ microwave powers, each of them respectively frequency-modulated at $\text{\SI{11}{\kilo\hertz}}$ and $\text{\SI{18}{\kilo\hertz}}$, with an amplitude of $\pm\text{\SI{5}{\kilo\hertz}}$. The power combination of these signals is, then, used to excite the GCPW, coupled to the resonators, in liquid He. After amplification and DC conversion at room temperature, by means of a low-noise amplifier and a diode detector, the AC component of the transmitted signal is detected by two different lock-in amplifiers, with a $\text{\SI{100}{\milli\second}}$ time constant, each of them referenced to the corresponding modulation frequency, in order to decouple the response of the individual resonators. Finally, the four signals, X and Y outputs for both resonators, are acquired by a DAQ with a $\text{\SI{100}{\milli\second}}$ integration time. The X-signals for both resonators are plotted in \autoref{PANEL_06:d}, showing a maximum in-resonance slope of about $\text{\SI{4.2}{\milli\volt/\mega\hertz}}$ for sensor $\alpha$ and $\text{\SI{8.9}{\milli\volt/\mega\hertz}}$ for sensor $\beta$. A $\text{\SI{2}{\hour}}$ time scan is then carried out in parallel on the two different devices, at their corresponding resonance frequencies. The measured voltages are converted in temperature variations by normalizing each dataset to the corresponding temperature responsivity (\textit{i.e.} $R_{\alpha}(\SI{4.2}{K}) = \text{\SI{4.3}{\milli\volt/\kelvin}}$ and $R_{\beta}(\SI{4.2}{K}) = \text{\SI{12.5}{\milli\volt/\kelvin}}$ respectively). Upon normalization, the responses of the two sensors show an almost identical time evolution [see \autoref{PANEL_06:e}], indicating that such low-frequency fluctuations might originate from a global common physical stimulus rather than from a noise source intrinsic to each sensor. Additionally, the coincidence of the two temperature estimations, originating from two sensors having significantly different responsivities, suggests the signal variations are associated to temperature-related phenomena, rather than to other physical quantities (\textit{i.e.} magnetic field), resulting from the cross-sensitivity of the superconducting sensor.

\begin{figure}[H]
\includegraphics[width=\linewidth]{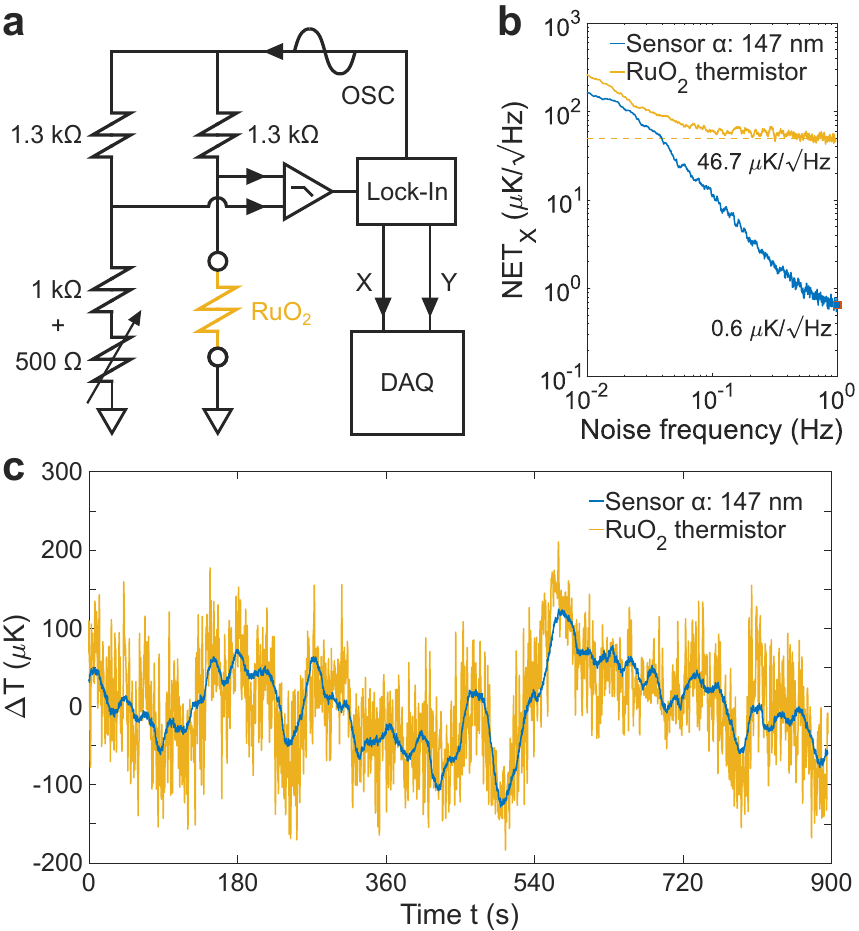}
\caption{\justifying Parallel temperature monitoring of the LHe bath at $\text{\SI{4.2}{\kelvin}}$, exploiting the $\text{\SI{147}{\nano\metre}}$ thick S-SRR and a commercial RuO\textsubscript{2} thermistor, excited in a lock-in detection configuration. a) AC Wheatstone bridge experimental setup used to bias the thermistor, when implementing a lock-in detection scheme. b) Frequency spectra of the $NET$ for the  $\text{\SI{147}{\nano\metre}}$ thick resonator (blue line) and the RuO\textsubscript{2} thermistor (yellow line). c) Simultaneous temperature monitoring of the low-frequency fluctuations, performed exploiting two lock-in detections on both the S-SRR sensor and the RuO\textsubscript{2} thermistor.}
\label{PANEL_07}
\makeatletter
\let\save@currentlabel\@currentlabel
\edef\@currentlabel{\save@currentlabel(a)}\label{PANEL_07:a}
\edef\@currentlabel{\save@currentlabel(b)}\label{PANEL_07:b}
\edef\@currentlabel{\save@currentlabel(c)}\label{PANEL_07:c}
\edef\@currentlabel{\save@currentlabel(b-c)}\label{PANEL_07:b-c}
\makeatother
\end{figure}

These hypotheses are confirmed by a final experiment, where the temperature inside the LHe dewar is simultaneously monitored by means of the $\text{\SI{147}{\nano\metre}}$ thick S-SRR and a commercial sensor relying on a different physical principle, in particular a RuO\textsubscript{2} thermistor \cite{Ihas:1998}. In particular, this device shows a resistance $R(\SI{4.2}{\kelvin}) = \text{\SI{1.36}{\kilo\ohm}}$ and a temperature sensitivity $dR/dT(\SI{4.2}{\kelvin}) = \text{\SI{-86}{\ohm/\kelvin}}$, for a $\text{\SI{10}{\mu\ampere}}$ biasing current. Considering the theoretical thermal noise limit associated to its resistance, with a $\sqrt{4k_BTR}/(I_{bias} \cdot dR/dT) \simeq \text{\SI{0.7}{\mu\kelvin/\hertz^{1/2}}}$ spectral plateau at $\text{\SI{4.2}{\kelvin}}$, such a thermistor should, in principle, allow for a sufficiently precise monitoring of the low-frequency temperature fluctuations around $\text{\SI{100}{\mu\kelvin}}$, showing a comparable resolution to the reported values for our S-SRRs. On the one hand, the excitation of the S-SRR is provided by the same instrumentation setup reported in \autoref{SECTION_A}, by acquiring the demodulated X-signal at the output of a lock-in amplifier: additionally, the excitation parameters are the same optimal ones exploited in the previous tests (\textit{i.e.} $P_{RF} = \text{\SI{-20}{\dBm}}$, $f_{FM} = \text{\SI{11}{\kilo\hertz}}$ and $\Delta f_{FM} = \pm\text{\SI{5}{\kilo\hertz}}$). On the other hand, the RuO\textsubscript{2} is biased by means of a room temperature Wheatstone bridge. A $\text{\SI{500}{\ohm}}$ potentiometer, in series to a $\text{\SI{1}{\kilo\ohm}}$ resistor, is placed on the branch opposite to the temperature-sensitive element, in order to accurately cancel out the differential voltage reading and balance the bridge. As described in the \textcolor{blue}{supplementary data, S6}, a simple DC bias of the bridge turns out to be insufficient for performing a temperature readout with a resolution in the order of $\text{\SI{100}{\mu\kelvin}}$, necessary to sense the low-frequency fluctuations under investigation. Therefore, as an attempt to minimize the noise associated to the RuO\textsubscript{2}, a lock-in detection scheme is implemented also on the temperature reading of the thermistor [see \autoref{PANEL_07:a}]. The oscillator output of a lock-in amplifier is exploited to bias the bridge with a $\text{\SI{26}{\milli\volt_{RMS}}}$ voltage, corresponding to a $\text{\SI{10}{\mu\ampere_{RMS}}}$ current flowing across the sensing branch. The biasing voltage is oscillating at $\text{\SI{17}{\kilo\hertz}}$, which is also exploited as lock-in internal modulation reference. The differential voltage reading is amplified, filtered (with a band-pass between $\text{\SI{1}{\kilo\hertz}}$ and $\text{\SI{100}{\kilo\hertz}}$) and acquired, as X-signal output of the lock-in amplifier, with a $\text{\SI{100}{\milli\second}}$ time constant, by a DAQ board with a $\text{\SI{100}{\milli\second}}$ integration time. A $\text{\SI{2}{\hour}}$ scan is performed, on both the X outputs related to the S-SRR and the RuO\textsubscript{2} sensor. In \autoref{PANEL_07:b-c} we report the results concerning a $\text{\SI{15}{\minute}}$ time lapse of the two temperature readings. Low-frequency fluctuations in the order of $\text{\SI{100}{\mu\kelvin}}$ are clearly distinguishable on both signals. The coincidence of the responses for the two sensors, each of them exploiting a different physical phenomenon\hfill to\hfill probe\hfill temperature,\hfill confirms\hfill the\hfill origin\hfill of 
\noindent such fluctuations to be related to environmental effects inducing sub-$\text{\SI{}{\milli\kelvin}}$ temperature variations in the LHe bath. Finally, the $NET$ spectra are extracted by applying an FFT on both the $\text{\SI{2}{\hour}}$ scan signals. While the $NET(\SI{1}{Hz})$ measured for the RuO\textsubscript{2} thermometer is about $\SI{50}{\mu\kelvin/\hertz^{1/2}}$, consistent with literature values \cite{Holmes:1992, Yeager:2001, Courts:2003, Ekin:2010}, our sensor offers a resolution of $\SI{0.6}{\mu\kelvin/\hertz^{1/2}}$, outperforming the thermistor almost by a factor $\text{100}$.

\section{Conclusion}
We have fabricated and tested a set of temperature-sensitive Nb\textsubscript{50}Ti\textsubscript{50} superconducting S-SRRs, generally showing $Q_L > \text{10\textsuperscript{4}}$ for $\SI{3}{\kelvin} < T < \SI{8}{\kelvin}$, in order to investigate the influence of the thin-film thickness on the resonance frequency shifts induced by the change of their kinetic inductance $L_{K}(T)$. In particular, an increase in the temperature sensitivity of the devices is observed, induced by larger kinetic inductance values for thinner films, with a maximum of $\text{\SI{19}{\mega\hertz/\kelvin}}$ for the $\text{\SI{24}{\nano\metre}}$ thick resonator. However, the significant reduction in the loaded quality factor, for thinner films, entails a reduction of the maximum slope, for the sensing DC and X curves. Indeed, the opposite behaviors of temperature sensitivity and Q-factor, for thinner films, induce an optimal thickness condition, when evaluating the temperature resolution of the devices. In this perspective, we have exploited frequency modulation and lock-in detection to assess the possibility to reach a sub-$\text{\SI{}{\mu\kelvin}}$ temperature resolution for such devices. Noise equivalent temperatures as low as $\text{\SI{0.5}{\mu\kelvin/\hertz^{1/2}}}$ are reported at $\text{\SI{1}{\hertz}}$ in liquid He for a $\text{\SI{105}{\nano\metre}}$ thick S-SRR. Additionally, the possibility to excite different resonators with the same microwave transmission line, while simultaneously and accurately tracking their temperature responses, has also been successfully demonstrated. We believe that such superconducting devices, eventually benefiting from higher kinetic inductance films (for enhanced sensitivity and further miniaturization) and enabling a frequency-multiplexed readout (for the parallel monitoring of different sensors), can find applications, directly integrated in the back-end of quantum architectures, in the localized monitoring of distributed on-chip cryogenic temperatures.

%
% Each of the commands below will create an unnumbered section with the appropriate heading.
% Remove any sections that are not relevant for your article.
% All sections except suppdata will be removed if the [anonymous] option is used.
% See iopjournal-guidelines.pdf for more information.
%

\ack{The authors thank the EPFL Center of Micro/Nanotechnology (CMi), in particular Cyrille Hibert, Julien Dorsaz, Niccolò Piacentini, Joffrey Pernollet and Adrien Toros, for the facility support provided during the microfabrication processing. We would also like to acknowledge Franco De Palma, Simone Frasca, Nergiz Sahin-Solmaz and Biranche Tandon for the fruitful discussions.}

\funding{This work has received funding from the Swiss National Science Foundation (SNSF), under the AMBIZIONE program (Project "Cryogenic Thermometry based on Superconducting Microwave Resonators", grant agreement No. PZ00P2\_193361).}
% This section is a list of funder names and grant numbers

\roles{
% List author names and the contributions made to the article, using terms from the NISO Contributor Roles Taxonomy (CRediT) https://credit.niso.org
\textbf{A. Chatel}: Data curation (lead); Formal analysis (lead); Investigation (lead); Methodology (lead); Software (lead); Visualization (lead); Writing - original draft (lead); Writing - review \& editing (equal);
\textbf{R. Russo}: Data curation (supporting); Formal analysis (supporting) ;Investigation (supporting); Methodology (equal); Software (equal); Writing - review \& editing (equal);
\textbf{L. Mazzone}: Formal analysis (supporting); Investigation (supporting); Methodology (supporting); Software (supporting); Writing - review \& editing (equal);
\textbf{Q. Boinay}: Formal analysis (supporting); Investigation (equal); Methodology (supporting); Writing - review \& editing (equal);
\textbf{R. Farsi}: Formal analysis (supporting); Investigation (supporting); Methodology (equal); Software (supporting); Writing - review \& editing (equal);
\textbf{J. Brügger}: Resources (equal); Supervision (equal); Validation (supporting); Writing - review \& editing (equal);
\textbf{G. Boero}: Conceptualization (equal); Data curation (equal); Formal analysis (equal); Funding acquisition (equal); Investigation (equal); Methodology (equal); Project administration (lead); Resources (equal); Software (equal); Supervision (lead); Validation (lead); Writing - original draft (equal); Writing - review \& editing (lead);
\textbf{H. Furci}: Conceptualization (lead); Data curation (equal); Formal analysis (equal); Funding acquisition (lead); Investigation (equal); Methodology (equal); Project administration (lead); Resources (lead); Software (equal); Supervision (lead); Validation (equal); Writing - original draft (equal); Writing - review \& editing (lead).
}

\data{The data that support the findings of this study are available upon reasonable request from the authors.}
% For more information on IOP Publishing's research data policy see: https://publishingsupport.iopscience.iop.org/questions/research-data/

\suppdata{The \textcolor{blue}{supplementary data} contains additional information about the following topics:
\vspace{-6pt}
\begin{enumerate}
\setlength{\itemsep}{-4pt}
\item[S1.] Comparative state-of-the-art on cryogenic thermometry;
\item[S2.] Microfabrication process flow and SEM/AFM characterization of the patterned S-SRRs;
\item[S3.] FEM and QUC microwave simulation of the resonance mode for the \SI{1.1}{\giga\hertz} S-SRR geometry;
\item[S4.] VTI temperature calibration of the S-SRRs;
\item[S5.] Determination of the modulation $P_{RF}$, $\Delta f_{FM}$ and $f_{FM}$ parameters in a LHe dewar;
\item[S6.] Simultaneous temperature monitoring in a LHe dewar, by means of an S-SRR and a RuO\textsubscript{2} thermistor.
\end{enumerate}}

\section*{Conflict of interest}
The authors have no conflicts to disclose.

\section*{Copyrights}
This article has been submitted for publication to IOP \textit{Superconductor Science and Technology}. Upon acceptance, it will be found at \url{https://iopscience.iop.org/journal/0953-2048}.

\section*{References}
\printbibliography[heading=none]

\end{multicols*}

\end{document}

\begin{center}
\LARGE{Effects of the thin-film thickness on superconducting NbTi microwave resonators for on-chip cryogenic thermometry: Supplementary data}\\
\vspace{4mm}
\Large{A. Chatel,\textsuperscript{1-2} R. Russo,\textsuperscript{1-2} L. Mazzone,\textsuperscript{1} Q. Boinay,\textsuperscript{1} R. Farsi,\textsuperscript{1} J. Brugger,\textsuperscript{1}\\ G. Boero\textsuperscript{1-2} and H. Furci\textsuperscript{1}}\\
\vspace{4mm}
\large{\textsuperscript{1}Microsystems Laboratory, EPFL, 1015 Lausanne, Switzerland}\\
\large{\textsuperscript{2}Center for Quantum Science and Engineeing, EPFL, 1015 Lausanne, Switzerland}\\
\end{center}
\vspace{4mm}
\normalsize

\section{Comparative state-of-the-art on cryogenic thermometry}
\label{SUPPL_MAT_00}

In this section we provide a more detailed description of the state-of-the-art on cryogenic thermometry. Even though developing a comprehensive review of all the current existing technologies lays beyond the scope of this work, we believe that such a brief overview is meaningful for the purpose of comparing the temperature performances of our resonators, highlighting advantages and drawbacks with respect to well-established commercial thermometers and research-level devices. \autoref{ADD_INFO_TABLE_00} reports a list of sensors that are currently widely used by the scientific community to monitor cryogenic temperatures.

\begin{table}[b!]
\caption{Different types of cryogenic temperature sensors, listed by technology, typical application, operating temperature range and resolution. The separation mid-rule is used to distinguish between commercial sensors, commonly used to monitor cryogenic temperatures on distributed stages of large scale environments, and superconducting-based research devices exploited to perform localized on-chip temperature sensing. Successive repeated entries in columns are indicated with a ditto mark ("). Missing data is indicated with a dash (-).\vspace{1mm}}
\label{ADD_INFO_TABLE_00}
\centering
\renewcommand{\arraystretch}{1.25}
\begin{threeparttable}
\begin{tabular*}{\columnwidth}{@{\hspace{10pt}\extracolsep{\fill}} ccccccc @{\hspace{10pt}}}
\toprule
Sensor & Technology & Typical use & T-range &  T-resolution\tnote{1} & Ref.\\
\midrule
Pt capsule & Resistor & Large scale & $\text{\SI{20}{\kelvin} - \SI{800}{\kelvin}}$ & - & \cite{Holmes:1992, Yeager:2001, Ekin:2010}\\
CERNOX\textsuperscript{\textregistered} film & " & " & $\text{\SI{100}{\milli\kelvin} - \SI{420}{\kelvin}}$ & $\text{\SI{100}{\mu\kelvin}}$ & \cite{Holmes:1992, Yeager:2001, Courts:2003, Ekin:2010, Courts:2014}\\
RuO\textsubscript{2} film & " & Large scale / On-chip & $\text{\SI{5}{\milli\kelvin} - \SI{300}{\kelvin}}$ & " &\cite{Holmes:1992, Yeager:2001, Ekin:2010, Wheeler:2020, Myers:2021}\\
Ge capsule & " & Large scale & $\text{\SI{50}{\milli\kelvin} - \SI{100}{\kelvin}}$ & " & \cite{Holmes:1992, Yeager:2001, Ekin:2010}\\
Si junctions& Diode & Large scale / On-chip & $\text{\SI{100}{\milli\kelvin} - \SI{500}{\kelvin}}$ & $\text{\SI{1}{\milli\kelvin}}$ & \cite{Holmes:1992, Yeager:2001, Ekin:2010, Courts:2016, Noah:2024}\\
GaAlAs junctions & " & Large scale & $\text{\SI{1}{\kelvin} - \SI{500}{\kelvin}}$ & $\text{\SI{100}{\mu\kelvin}}$ & \cite{Holmes:1992, Yeager:2001, Ekin:2010}\\
Au–Fe wire & Thermocouple & " & $\text{\SI{1}{\kelvin} - \SI{1500}{\kelvin}}$ & $\text{\SI{10}{\milli\kelvin}}$ & \cite{Holmes:1992, Ekin:2010}\\
Capacitance sensor & Capacitor & " & $\text{\SI{1}{\kelvin} - \SI{300}{\kelvin}}$ & $\text{\SI{1}{\milli\kelvin}}$ & \cite{Holmes:1992, Ekin:2010}\\
\midrule
$I_C(T)$ tunnel-junction & DC Superconductor & On-chip & $\text{\SI{1}{\milli\kelvin} - \SI{100}{\milli\kelvin}}$ & - & \cite{Feshchenko:2015}\\
$I_C(T)$ thin film & " & " & $\text{\SI{600}{\milli\kelvin} - \SI{1.2}{\kelvin}}$ & \textcolor{white}{\tnote{2}}-\tnote{2} & \cite{Noah:2024}\\
TLS resonator & RF Superconductor & " & $\text{\SI{50}{\milli\kelvin} - \SI{1}{\kelvin}}$ & \textcolor{white}{\tnote{3}}-\tnote{3} & \cite{Wheeler:2020}\\
$L_K(T)$ resonator & " & " & $\text{\SI{3}{K} - \SI{8}{\kelvin}}$ & $\text{\SI{500}{\nano\kelvin}}$ & Our work\\
\bottomrule
\end{tabular*}
\begin{tablenotes}
\item[1] Values estimated, for a $\text{\SI{1}{\hertz}}$ bandwidth, at $\text{\SI{1}{\hertz}}$ and at a $\text{\SI{4.2}{\kelvin}}$ operational temperature.
\item[2] $\text{\SI{50}{}-\SI{200}{\milli\kelvin}}$ temperature resolution achieved in the specific range.
\item[3] $\text{\SI{50}{}-\SI{75}{\mu\kelvin}}$ temperature resolution achieved in the specific range.
\end{tablenotes}
\end{threeparttable}
\end{table}

As a matter of  fact, many cryogenic thermometers are nowadays commercially available, exploiting a whole variety of technologies\cite{Holmes:1992, Yeager:2001, Courts:2003, Ekin:2010}. Among them, variable-temperature resistors (\textit{a.k.a.} thermistors), in the form of Zirconium-oxy-nitride (often referred by the commercial name of CERNOX\textsuperscript{\textregistered})\cite{Courts:2003} and Ruthenium-oxide\cite{Myers:2021} films, represent the market-leaders. Thanks to their mechanical robustness, ease of signal measurement, wide operational range (\textit{i.e.} $\text{\SI{1}{}-\SI{300}{\kelvin}}$), good temperature resolution (\textit{i.e.} $\text{\SI{0.1}{}-\SI{100}{\milli\kelvin}}$) and low sensitivity to magnetic fields (especially for the RuO\textsubscript{2} sensors), thermistors are successfully employed in common cryogenic apparatuses, where many temperature stages, as well as distributed locations, are required to be simultaneously monitored on a large scale.

However, the recent advancements on miniaturized systems, for instance, for quantum computing and quantum sensing, are requiring innovative paradigms for locally monitoring cryogenic temperatures directly on-chip. As a matter of fact, thermistors are typically bulky devices, whose dimensions, power consumption and need for individual electrical routing introduce serious design limitations when allocating multiple sensors at the chip level (\textit{i.e.} for integration areas smaller than $\text{\SI{100}{\milli\metre^2}}$). The possibility to exploit superconducting materials, benefiting from extremely low power consumptions and well-established micro-fabrication techniques, has been successfully proven to represent a valid alternative for localized on-chip temperature monitoring\cite{Feshchenko:2015, Wheeler:2020, Noah:2024}. In particular, Wheeler et al.\cite{Wheeler:2020} have successfully demonstrated the possibility to exploit the temperature-dependence of two-levels systems (TLSs) microwave resonator to accurately track the self-heating of their kinetic inductance traveling-wave parametric amplifier (KIT). Their devices, directly anchored to the top surface of the KIT, have been able to reach temperature resolutions of about $\text{\SI{50}{\mu\kelvin/\hertz^{1/2}}}$, below $\text{\SI{1}{\kelvin}}$ (where TLS represents the major responsible for the resonance frequency shifts). More recently, Noah et al.\cite{Noah:2024} have presented a set of four temperature-sensitive devices, intrinsically integrated in the CMOS-compatible stack of their quantum chip. Among them, the temperature-dependence of the DC critical current  $I_C(T)$ of a superconducting adhesion layer has been exploited to monitor on-chip temperatures with a resolution as low as $\text{\SI{50}{\milli\kelvin/\hertz^{1/2}}}$, below the critical transition temperature of $\text{\SI{1.2}{\kelvin}}$.

The thin-film superconducting resonators presented in our work, based on the kinetic inductance $L_K(T)$-induced resonance frequency shifts, show temperature resolutions of about $\text{\SI{500}{\nano\kelvin/\hertz^{1/2}}}$ at $\text{\SI{1}{\hertz}}$ and $\text{\SI{4.2}{\kelvin}}$, improving this figure-of-merit (FOM) by a factor $\text{\SI{100}{}}$. Even though superconducting materials are limited in the operational temperature range and show a non-negligible sensitivity to magnetic fields, the possibility to simultaneously excite several devices by the same microwave coplanar line, their high Q-factor (with an intrinsically low power dissipation), the ease of integration (eventually, directly at the chip back-end level) and miniaturization could make them promising candidates for localized on-chip thermometry of complex quantum systems and architectures.\\
Nevertheless, several crucial aspects have to be carefully considered in the design phase, in order to optimally exploit such devices for a high-resolution temperature sensing. First of all, the superconducting critical temperature $T_C$ plays a key role in the choice of the material, in order to ensure compatibility with the operating temperature of the monitored chip. In such a context, Al can be exploited, for instance, to perform temperature readings below $\text{\SI{1}{\kelvin}}$ (suitable for working in dilution refrigerators), while high-temperature superconductors, like YBCO, are ideal for liquid nitrogen applications (\textit{i.e.} for $T > \text{{\SI{77}{\kelvin}}}$). Second, superconducting materials are sensitive to magnetic fields, whose presence can significantly modify the behavior of the resonators (\textit{i.e.} by decreasing $T_C$ and increasing the slope of the $f_{res}$ VS $T$ response curve) and introduce an additional contribution to the signal noise. Since quantum systems are typically operated either in magnetically shielded conditions, or by applying external DC fields, uniform over the chip plane, superconducting $L_K(T)$-based resonators would require a preliminary calibration against magnetic fields, in order to properly estimate cryogenic temperature variations. Finally, a further miniaturization of such resonators has to ultimately consider the $\propto f^2$ increase of the superconducting surface resistance. Indeed, for smaller dimensions, the geometric capacitance and inductance can drastically reduce, increasing the design resonance frequency and significantly lowering down the intrinsic quality factor of such devices. In order to still apply a favorable scaling law, materials with higher kinetic inductance, as TiN, NbN and NbTiN, could be exploited for counterbalancing the increase of the operating resonance frequency.

\section{Microfabrication of the superconducting S-SRRs}
\label{SUPPL_MAT_01}

\begin{figure*}[t!]
\includegraphics[width=\textwidth]{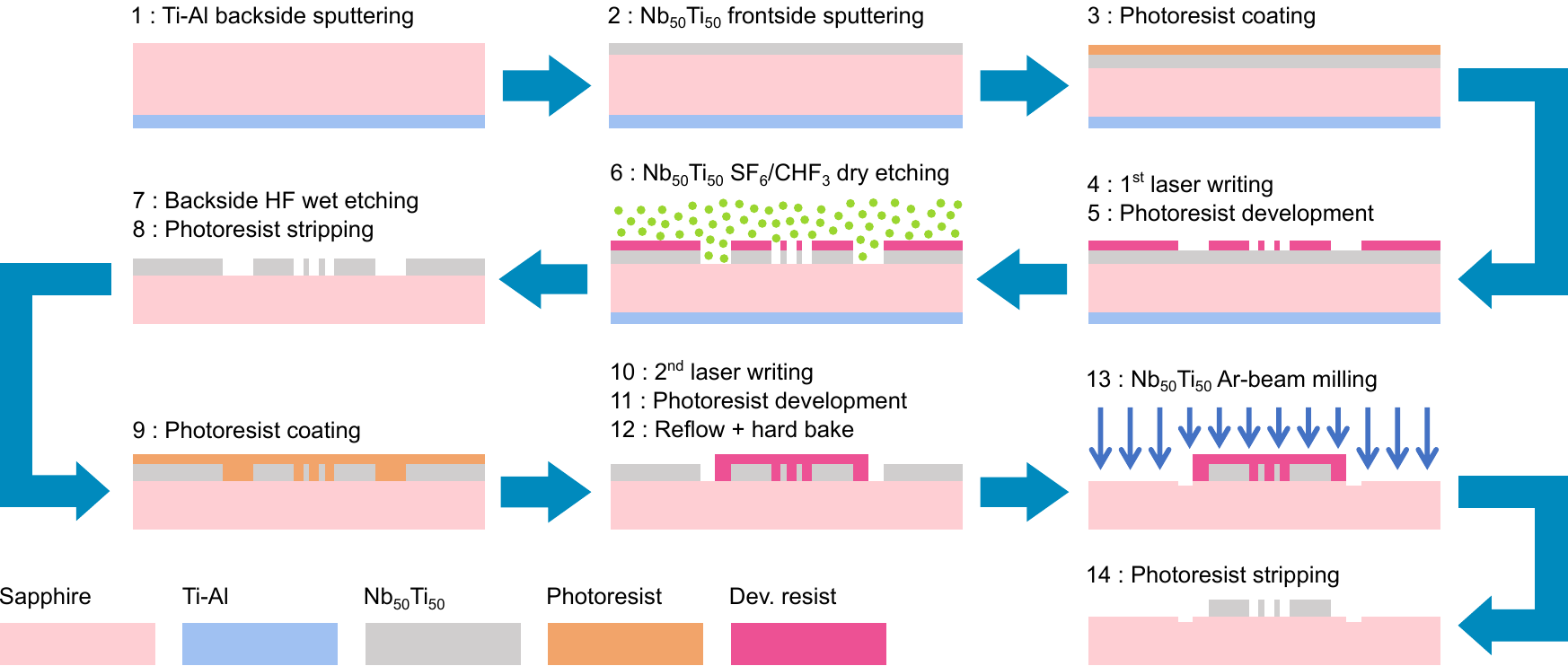}
\caption{Microfabrication process flow to pattern Nb\textsubscript{50}Ti\textsubscript{50} thin-films on top of sapphire wafers.}
\label{ADD_INFO_01_a}
\end{figure*}

This section is dedicated to the description of the microfabrication of the  S-shaped split ring resonators (S-SRRs). In \autoref{ADD_INFO_01_a} we schematize the process flow to pattern Nb\textsubscript{50}Ti\textsubscript{50} thin-films (with thickness values approximatively ranging from $\text{\SI{20}{\nano\metre}}$ to $\text{\SI{150}{\nano\metre}}$) on top of $\text{\SI{650}{\mu\metre}}$ thick C-plane sapphire wafers. The first two steps involve the deposition of a backside $\text{\SI{10}{\nano\metre}}$-$\text{\SI{100}{\nano\metre}}$ Ti-Al bilayer (necessary to perform an electrostatic wafer clamping during the plasma etching process) and the superconducting Nb\textsubscript{50}Ti\textsubscript{50} front layer. In both cases, the materials are DC sputtered on the sapphire substrate, kept at room temperature. Two consecutive runs of direct laser writing photolithography ($\lambda = \text{\SI{355}{\nano\metre}}$) followed by two different dry etching processes are, then, performed in order to pattern the S-SRR resonators. In particular, the first high-resolution process exploits a fluorine-based plasma etching to pattern the S-SRR geometries in small squared openings. As a matter of fact, when the design pattern consists of large regions exposed to plasma, electrostatic de-clamp of the wafer frequently occurs, attributed to a significant deformation of the sapphire substrate, probably due to the thin-film stress induced by a large temperature increase during the etching process. This issue is avoided when the area of the open regions is limited to the minimum around the highly resolved resonator patterns. The second lithography-etching run, relying on the use of Ar-ion beam milling, is necessary to remove the remaining metallic layer. Indeed, the edge-clamping mechanism of our milling machine is exclusively mechanical, preventing any wafer de-clamp due to thermal stress and overheating. The use of the plasma etching over this latter technique is preferred due to the loss of resolution and wall sharpness induced by the thermal reflow of the photoresist. In the end, the wafer is diced into $\text{\SI{12.5}{\milli\metre}}$ $\times$ $\text{\SI{8.5}{\milli\metre}}$ rectangular chips, each containing two different S-SRRs. The presence of two resonators, showing a similar geometry but a significant difference in overall sizes, has to be attributed to our initial intention to perform a simultaneous frequency-multiplexed readout in different regions of the microwave spectrum.

\begin{figure*}[t!]
\includegraphics[width=\textwidth]{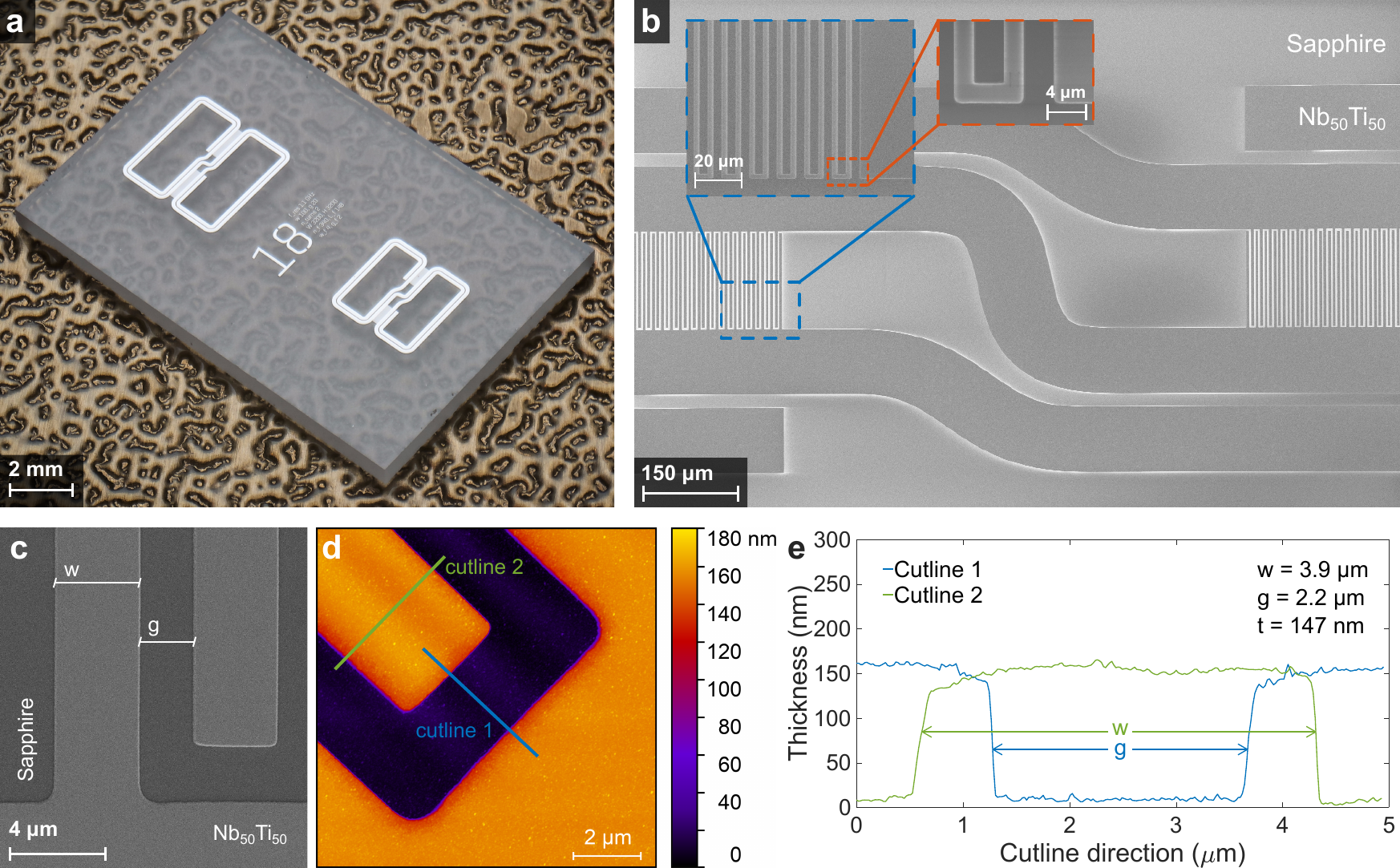}
\caption{Characterization of the microfabrication process flow. a) Optical microscope picture of the sapphire chip: the DUT corresponds to the top-left resonator. b) Set of SEM images, zooming on the interdigitated fingers region of the S-SRR. c) SEM picture highlighting some fingers of the IDC. d) AFM zoomed scan of the previous structure: the Nb\textsubscript{50}Ti\textsubscript{50} film is supposed to be $\text{\SI{150}{\nano\metre}}$ thick by design. e) AFM cross-section cuts of a single finger, along its width and gap size.}
\label{ADD_INFO_01_b}
\makeatletter
\let\save@currentlabel\@currentlabel
\edef\@currentlabel{\save@currentlabel(a)}\label{ADD_INFO_01_b:a}
\edef\@currentlabel{\save@currentlabel(b)}\label{ADD_INFO_01_b:b}
\edef\@currentlabel{\save@currentlabel(c)}\label{ADD_INFO_01_b:c}
\edef\@currentlabel{\save@currentlabel(d)}\label{ADD_INFO_01_b:d}
\edef\@currentlabel{\save@currentlabel(e)}\label{ADD_INFO_01_b:e}
\edef\@currentlabel{\save@currentlabel(b-c)}\label{ADD_INFO_01_b:b-c}
\edef\@currentlabel{\save@currentlabel(d-e)}\label{ADD_INFO_01_b:d-e}
\makeatother
\end{figure*}

\autoref{ADD_INFO_01_b} reports the main results concerning the patterning quality of the fabricated devices. Each diced chip contains two resonators [see \autoref{ADD_INFO_01_b:a}], the smallest and the largest ones designed to respectively resonate around $\text{\SI{1.6}{\giga\hertz}}$ and $\text{\SI{1.1}{\giga\hertz}}$. Considering the S-SRR geometry of this latter, being the only device under test (DUT) analyzed all over this work, the critical dimensions of such a design are related to the interdigitated finger capacitors (finger length :$\text{\SI{148}{\mu\metre}}$, finger width: $\text{\SI{4}{\mu\metre}}$, inter-finger gap: $\text{\SI{2}{\mu\metre}}$). Even though scanning electron microscopy (SEM) pictures allow to assess the quality of the overall microfabrication, confirming the absence of major defects [see \autoref{ADD_INFO_01_b:b-c}], atomic force microscopy (AFM) scans quantitatively estimate the pattern accuracy. In particular, from \autoref{ADD_INFO_01_b:d-e} it is possible to determine the design transfer inaccuracy of the patterned features, lower than $\text{\SI{300}{\nano\metre}}$, as well as the exact film thickness attained through the Nb\textsubscript{50}Ti\textsubscript{50} thin-film sputtering. Similar measurements are carried out on each of the samples fabricated for this work.

\section{Microwave simulation of the GCPW-coupled S-SRRs}
\label{SUPPL_MAT_02}

The aim of this section is to present the electromagnetic simulation of the proposed S-SRR geometry. 

\begin{figure*}[t!]
\includegraphics[width=\textwidth]{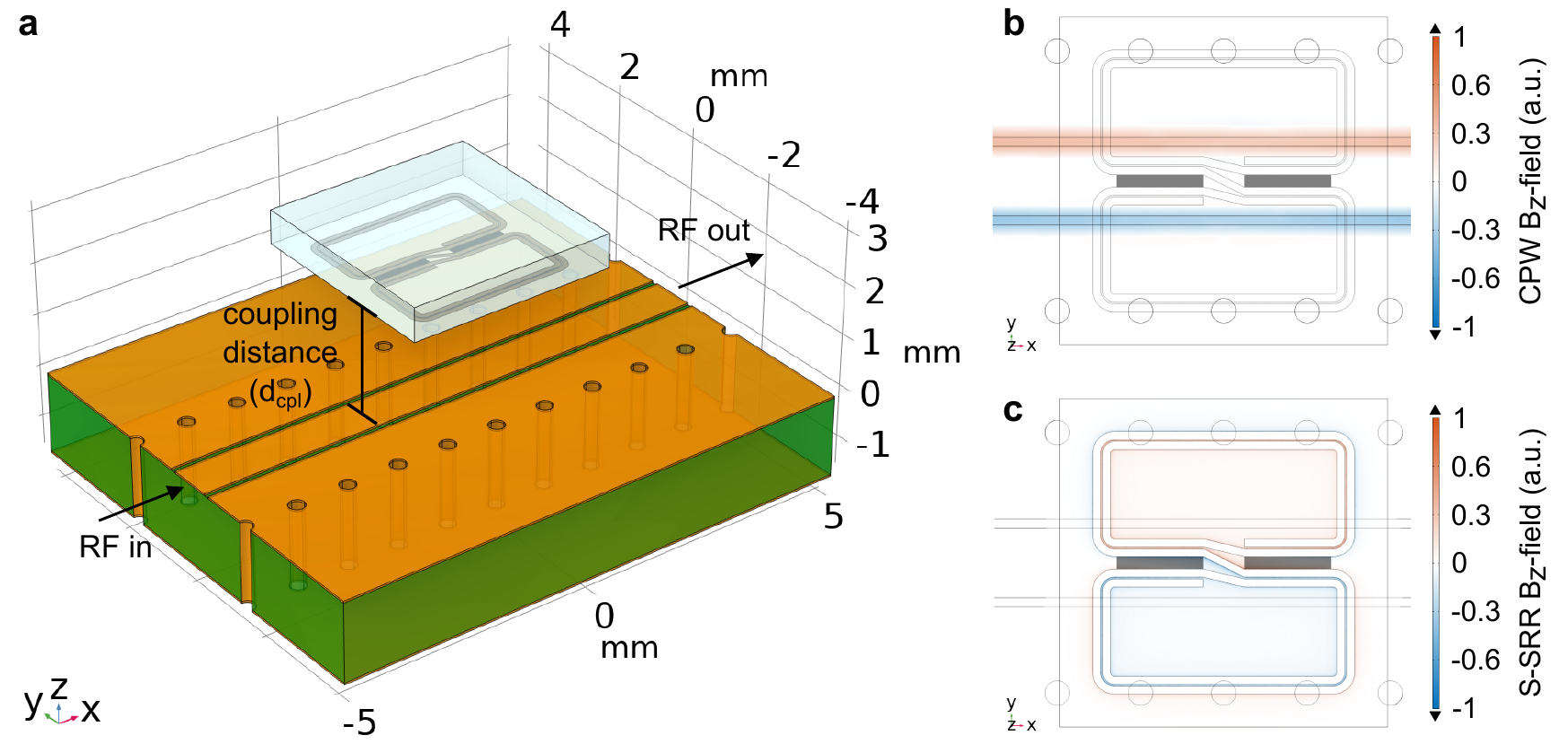}
\caption{COMSOL Multiphysics\textsuperscript{\textregistered} electromagnetic microwave simulation of the S-SRR excited by a GCPW. a) Schematic presenting the simulated 2-port geometry. The GCPW consists of a $\text{\SI{34}{\mu\metre}}$ thick copper plate on top of a $\text{\SI{1.5}{\milli\metre}}$ thick ROGERS 4003C\texttrademark{} laminate ($\epsilon_r \simeq 3.38$ and $\tan\delta \simeq 0.0027$). The central signal line is $\text{\SI{850}{\mu\metre}}$ wide and $\text{\SI{110}{\mu\metre}}$ laterally spaced from the ground plane. b-c) Antisymmetric magnetic field distribution of the GCPW, exciting the characteristic resonance S-SRR mode.}
\label{ADD_INFO_02_a}
\makeatletter
\let\save@currentlabel\@currentlabel
\edef\@currentlabel{\save@currentlabel(a)}\label{ADD_INFO_02_a:a}
\edef\@currentlabel{\save@currentlabel(b)}\label{ADD_INFO_02_a:b}
\edef\@currentlabel{\save@currentlabel(c)}\label{ADD_INFO_02_a:c}
\edef\@currentlabel{\save@currentlabel(b-c)}\label{ADD_INFO_02_a:b-c}
\makeatother
\end{figure*}

COMSOL Multiphysics\textsuperscript{\textregistered} and, specifically, its \textit{Electromagnetic Waves, Frequency Domain (emw)} environment are first exploited to estimate the resonance frequency of the thin-film microresonator, coupled to the microwave grounded coplanar waveguide (GCPW). \autoref{ADD_INFO_02_a:a} presents the schematic of the simulated 2-port geometry. A microwave power is injected from port 1 on the left and collected at port 2 on the right. The resonator is placed face-down at a $\text{\SI{2.5}{\milli\metre}}$ coupling distance $d_{cpl}$. Because of the extremely low aspect ratio of the device, the Nb\textsubscript{50}Ti\textsubscript{50} film is simulated as a bi-dimensional perfect electric conductor (PEC). This simplification does not allow to study the superconducting RF properties of the film, but it results to be sufficiently accurate and computationally efficient to estimate the resonance frequency of the device to be around $\text{\SI{1.1}{\giga\hertz}}$. For such a geometry, the total geometric inductance is estimated by analytical expressions\cite{Mohan:1999, Aebischer:2020} as $L_G\sim\text{\SI{39}{\nano\henry}}$. By knowing the resonance frequency value from simulation, the total geometric capacitance is calculated as $C_G = 1/\bigl((2\pi f)^2L_G\bigr) \sim\text{\SI{570}{\femto\farad}}$. The GCPW port-matching is also evaluated for the propagation quasi-TEM mode, getting a characteristic impedance $Z_{C} \text{ = \SI{48}{\ohm}}$. The characteristic S-SRR eigenmode is properly excited by the GCPW antisymmetric magnetic field configuration, allowing the resonator to inductively couple to the transmission line. [see \autoref{ADD_INFO_02_a:b-c}].

\begin{figure*}[t!]
\includegraphics[width=\textwidth]{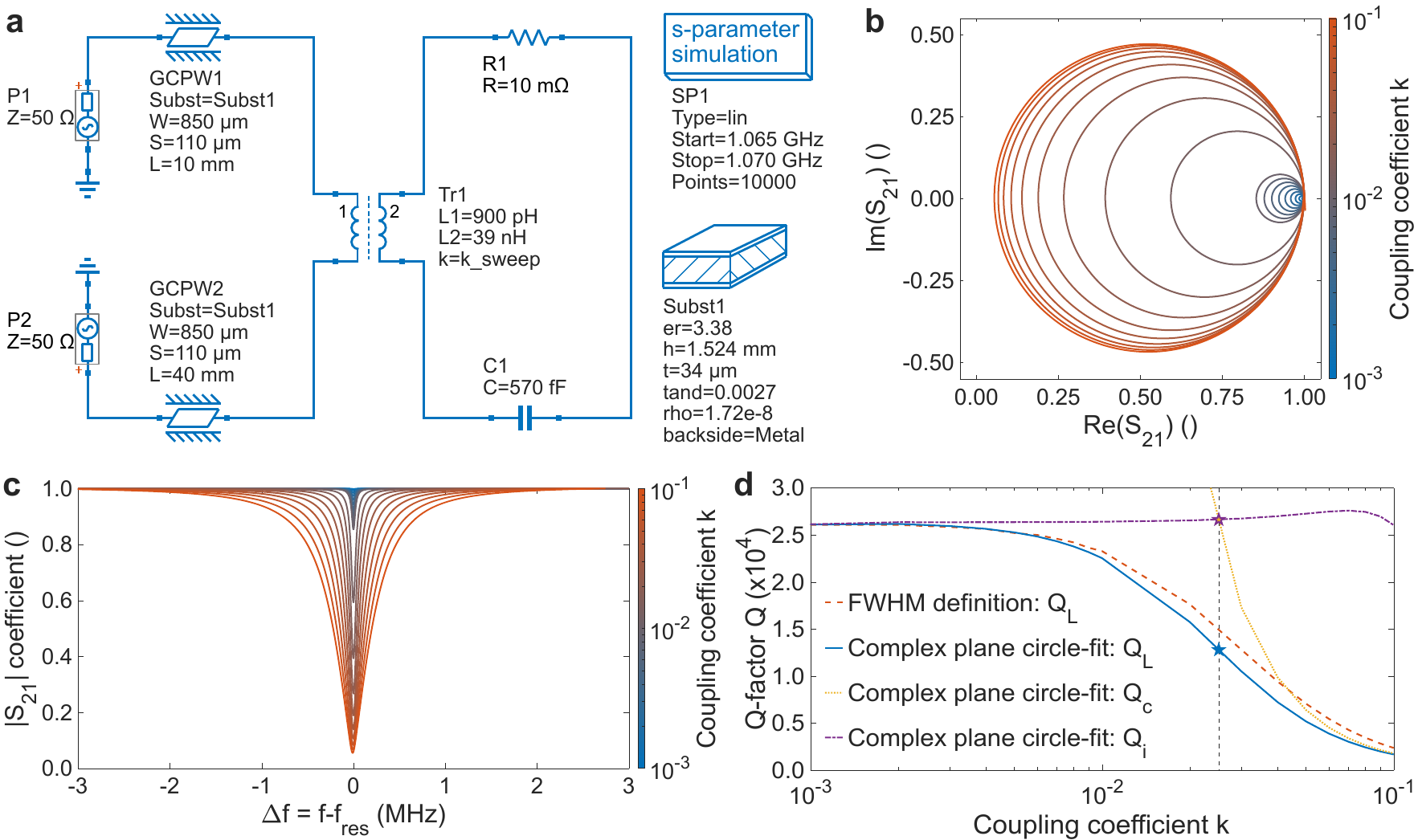}
\caption{2-ports microwave circuit simulation of the RLC resonator magnetically coupled to a coplanar line. a) QUCS schematic highlighting the exciting GCPW and RLC coupled components. b) Evolution of the real and imaginary parts of the $S_{21}$ parameter with respect to the magnetic coupling coefficient $k$. The complex data are fitted to the notch-type model and, then, renormalized to the obtained $ae^{i(\alpha-2\pi\tau f)}$ term, to compensate for the environmental effects. d) Saturation of the Q-factor, in the weak coupling limit, to the internal  $Q_i$ of an isolated RLC resonator. The single pentagram points represent the critical coupling condition (\textit{i.e.} $Q_i = Q_c \rightarrow Q_L = Q_i/2$ ).}
\label{ADD_INFO_02_b}
\makeatletter
\let\save@currentlabel\@currentlabel
\edef\@currentlabel{\save@currentlabel(a)}\label{ADD_INFO_02_b:a}
\edef\@currentlabel{\save@currentlabel(b)}\label{ADD_INFO_02_b:b}
\edef\@currentlabel{\save@currentlabel(c)}\label{ADD_INFO_02_b:c}
\edef\@currentlabel{\save@currentlabel(d)}\label{ADD_INFO_02_b:d}
\edef\@currentlabel{\save@currentlabel(b-c)}\label{ADD_INFO_02_b:b-c}
\makeatother
\end{figure*}

The dependence of the transmission resonance peak on the electromagnetic coupling coefficient $k$ is analyzed, in a second time, by means of a Quite Universal Circuit Simulator (QUCS) $k$-sweep. As shown in \autoref{ADD_INFO_02_b:a}, a 2-ports signal excitation of the coplanar line (whose geometric parameters coincide with the ones listed in \autoref{SUPPL_MAT_02}) is exploited to magnetically couple the line inductance $L_1 = \text{\SI{900}{\pico\henry}}$ to the resonator inductance $L_2 = \text{\SI{39}{\nano\henry}}$. The resonator is modeled as an RLC circuit, where $C = \text{\SI{570}{\femto\farad}}$ and $R = \text{\SI{10}{\milli\ohm}}$ (to match a limiting estimation of the unloaded Q-factor around $\text{\SI{2.60}{}$\times$10\textsuperscript{4}}$). As described in several studies\cite{Gao:2008, Khalil:2012, Probst:2015}, the typical $S_{21}$ transmission response of a notch-type resonator can be defined as:
\begin{equation}
\label{EQUATION_01}
S_{21}(f) = a e^{i(\alpha-2\pi\tau f)}\biggl( 1-\frac{(Q_L/|Q_c|)e^{i\phi}}{1+2iQ_L(f/f_{res}-1)} \biggr)
\end{equation}
where the term $a e^{i(\alpha-2\pi\tau f)}$ accounts for the environmental effects introduced by the excitation microwave cabling (\textit{i.e.} $a$ for the amplitude attenuation, $\alpha$ for the phase shift and $\tau$ for the line delay) and $\phi$ takes into consideration impedance mismatches at the input/output ports. Moreover, $f_{res}$ represents the resonance frequency of the RLC device, while $Q_L$ and $Q_c$ denote, respectively, the loaded and coupling quality factors. These latter are linked, together with the $Q_i$ internal Q-factor of the resonator, by the constitutive relation $Q_L^{-1} = Q_i^{-1} + Q_c^{-1}$. The results of such S-parameter simulation, run over a $k$-sweep from 10\textsuperscript{-3} to 10\textsuperscript{-1}, are presented in \autoref{ADD_INFO_02_b:b-c}, showing the evolution of a resonance peak located around $\text{\SI{1.0675}{\giga\hertz}}$. Upon renormalizing the complex $S_{21}$ data to the environmental pre-factor and subsequently fitting to \autoref{EQUATION_01}, the dependence of the $Q_L$, $Q_c$ and $Q_i$ quality factors on the coupling coefficient is shown in \autoref{ADD_INFO_02_b:d}, highlighting the theoretical saturation to the unloaded case $Q_i = 2 \pi f_{res} L/R$, for the weak coupling limit.

\section{Temperature calibration of the S-SRRs}
\label{SUPPL_MAT_03}

\begin{figure*}[ht!]
\includegraphics[width=\textwidth]{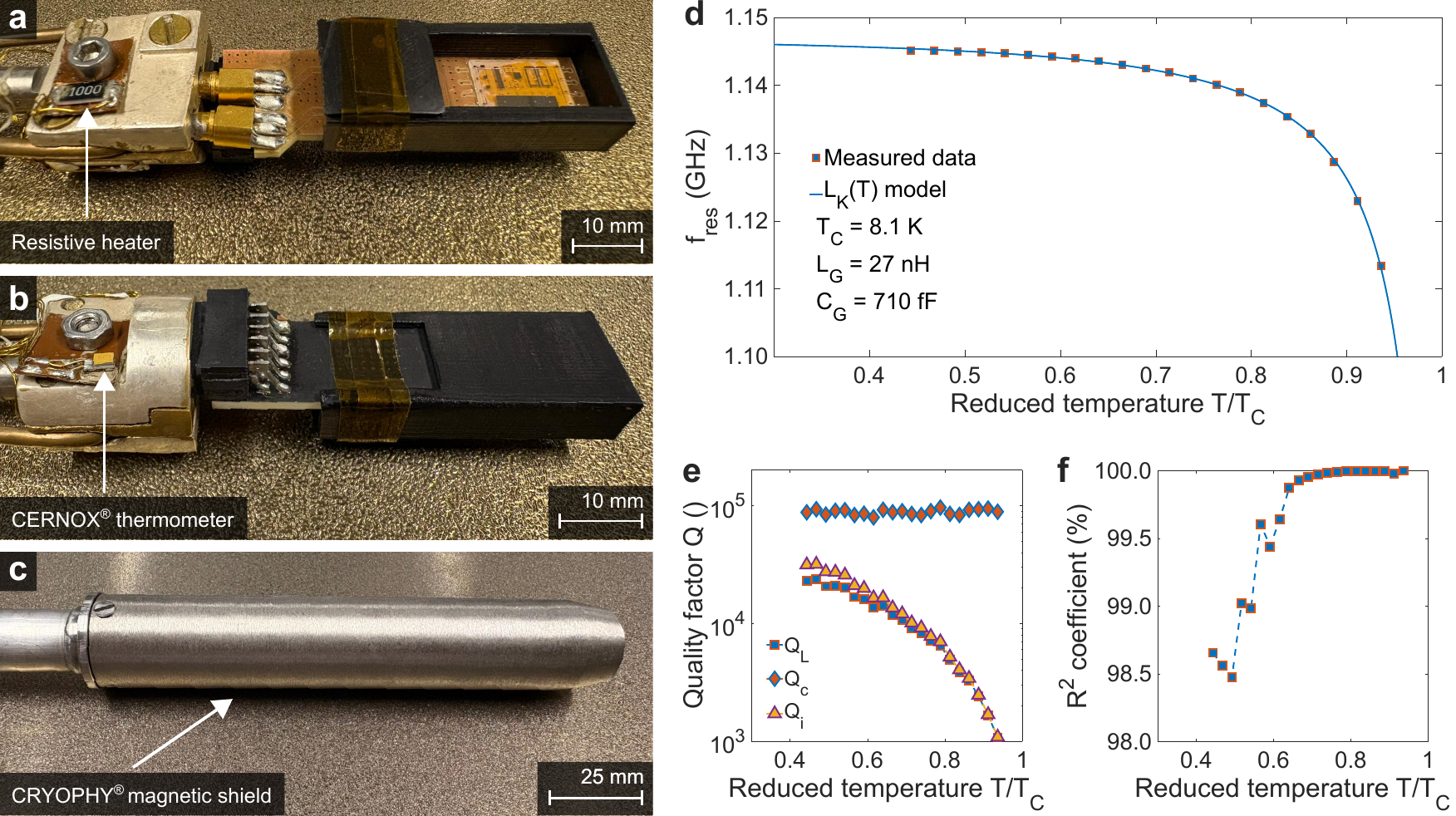}
\caption{Temperature calibration of the superconducting S-SRRs. a) Front side of the PCB, connected to the cryogenic probe. The resistive heater, exploited for controlling the temperature, is shown on the left. b) Back side of the PCB, connected to the cryogenic probe. The CERNOX\textsuperscript{\textregistered} thermometer, exploited for measuring the chip temperature, is shown on the left. c) Bottom extremity of the cryogenic probe, enclosed by the CRYOPHY\textsuperscript{\textregistered} magnetic shield. d) Typical temperature evolution of the resonance frequency, for the $\text{\SI{147}{\nano\metre}}$ thick resonator. The squared points represent the measured data, while the continuous line stands for the $L_K(T)$ model. e) Temperature dependence of the loaded $Q_L$, coupling $Q_c$ and internal $Q_i$ quality factors. f) $R^2$ notch-type resonator complex plane fitting coefficient, for each frequency scan at different temperatures.}
\label{ADD_INFO_03}
\makeatletter
\let\save@currentlabel\@currentlabel
\edef\@currentlabel{\save@currentlabel(a)}\label{ADD_INFO_03:a}
\edef\@currentlabel{\save@currentlabel(b)}\label{ADD_INFO_03:b}
\edef\@currentlabel{\save@currentlabel(c)}\label{ADD_INFO_03:c}
\edef\@currentlabel{\save@currentlabel(d)}\label{ADD_INFO_03:d}
\edef\@currentlabel{\save@currentlabel(e)}\label{ADD_INFO_03:e}
\edef\@currentlabel{\save@currentlabel(f)}\label{ADD_INFO_03:f}
\edef\@currentlabel{\save@currentlabel(a-b)}\label{ADD_INFO_03:a-b}
\makeatother
\end{figure*}

In order to provide a proper estimation of the temperature sensitivity below the critical superconducting transition, as well as an accurate reading of the cryogenic temperature, the S-SRRs have to be preliminarily calibrated against another type of thermometer. As a matter of fact, the $L_K(T)$-induced frequency shifts of a superconducting resonator represent a type of secondary thermometry, differentiating from primary thermometry, involving only fundamental physical laws and constants. The aim of this section is, therefore, to present the procedure exploited to perform the temperature calibration of the S-SRRs against a set of commercially available CERNOX\textsuperscript{\textregistered} thermistors. The temperature monitoring and control is achieved through the use of two CERNOX\textsuperscript{\textregistered} thermometers and two heaters. The thermometer-heater couple closest to the microwave resonator is mounted at about $\text{\SI{20}{\milli\metre}}$ from the two microwave connectors on the PCB [see \autoref{ADD_INFO_03:a-b}]. The ground electrical connection of these latter is in thermal contact with a 4 cm$^3$ copper block on which the thermometer and the heater are mounted. The other couple is mounted on the bottom of the VTI and is used to control the temperature of the He gas flowing around the resonator. Typically this temperature is set $\text{\SI{0.2}{\kelvin}}$ below the desired one in proximity of the device and measured with the other thermometer.\\
Moreover, the front side of the PCB is connected, in transmission, to two BeCu coaxial cables, used to excite the superconducting sensor placed $\text{\SI{2.5}{\milli\metre}}$ far from the surface, while its back side is dedicated to the DC routing to perform the cryogenic characterization of the thin-film. Additionally, as shown in \autoref{ADD_INFO_03:c}, in order to reduce the influence of external magnetic fields, the whole system, comprising of the sensing S-SRR and excitation PCB, is encapsulated in a cylindrical magnetic shield (from \textit{Magnetic Shields Ltd}) made of CRYOPHY\textsuperscript{\textregistered}, a Ni-Fe soft magnetic alloy typically exploited to shield magnetic fields in cryogenic environments ($\mu_r$ around $ \text{\SI{7}{}$\times$10\textsuperscript{4}}$ at $\text{\SI{4}{\kelvin}}$).

\autoref{ADD_INFO_03:d} presents the evolution of the resonance frequency of a single S-SRR, $\text{\SI{147}{\nano\metre}}$ thick. The Cu GCPW is excited through a VNA (\textit{PicoVNA 108}), directly connected in transmission to the PCB by BeCu coaxial cables, delivering a microwave power of $\text{\SI{-18}{dBm}}$. Frequency scans are collected at different temperatures and, then, fitted to the notch-type resonator complex plane model\cite{Probst:2015} [see \autoref{EQUATION_01}], for extracting the resonance frequency and the loaded, coupling and internal quality factor values. The temperature variation of $f_{res}$ is additionally fitted to a kinetic inductance-based model\cite{Tinkham:2004, Annunziata:2010, Nulens:2023}, showing a good agreement with the acquired data and the estimation of the $T_C$, $L_G$ and $C_G$ fitting parameters. In \autoref{ADD_INFO_03:e} we report the temperature dependence of the loaded $Q_L$, coupling $Q_c$ and internal $Q_i$ quality factors, estimated notch-type resonator complex plane model. The trends follow the physical behaviour of the system, with $Q_c$ maintaining a constant value of about $10^5$ and $Q_i$ decreasing when approaching the critical temperature $T_C$, due to the decrease of the Cooper pairs density $n_S(T)$ and the consequent increase in the microwave resistance of the superconducting thin-film. The goodness of the resonance peak fitting is also demonstrated by the $R^2$ coefficient, larger than $\text{99.5\%}$ for the majority of the frequency scans [see \autoref{ADD_INFO_03:f}]. We suspect such a parameter to slightly degrade for $T/T_C < 0.7$ in reason of the larger Q-factor, which translates into a lower number of peak points available for fitting (considering a fixed frequency scan window).

\section{Determination of the modulation parameters}
\label{SUPPL_MAT_04}

Even though a VNA-based measurement scheme (\textit{PicoVNA 108}) is preferentially exploited when performing fast and highly refined frequency scans (\textit{i.e} more than 1000 points, with a $\text{\SI{10}{\kilo\hertz}}$ acquisition bandwidth, as in \autoref{SUPPL_MAT_04}), a preliminary test has been carried out in LHe, in order to estimate the noise equivalent temperature ($NET$) of the S-SRRs at $\text{\SI{4.2}{\kelvin}}$. Upon signal-to-temperature conversion, $NET$ values always larger than $\text{\SI{20}{\mu\kelvin/\hertz^{1/2}}}$ have been recorded at $\text{\SI{1}{\hertz}}$, from a fast Fourier transform (FFT) applied on $\text{\SI{2}{\hour}}$ long time scans. As a consequence, a more sophisticated setup, based on lock-in signal detection, is implemented, in an attempt to further improve such a FOM by applying frequency modulation (FM) techniques. This section is, therefore, dedicated to the description of the instrumentation used to characterize the $NET$ of the resonators, in liquid He at $\text{\SI{4.2}{\kelvin}}$, as well as the preliminary tests carried out to identify the optimal modulation parameters for performing such an estimation of the temperature resolution.

\begin{figure*}[h!]
\includegraphics[width=\textwidth]{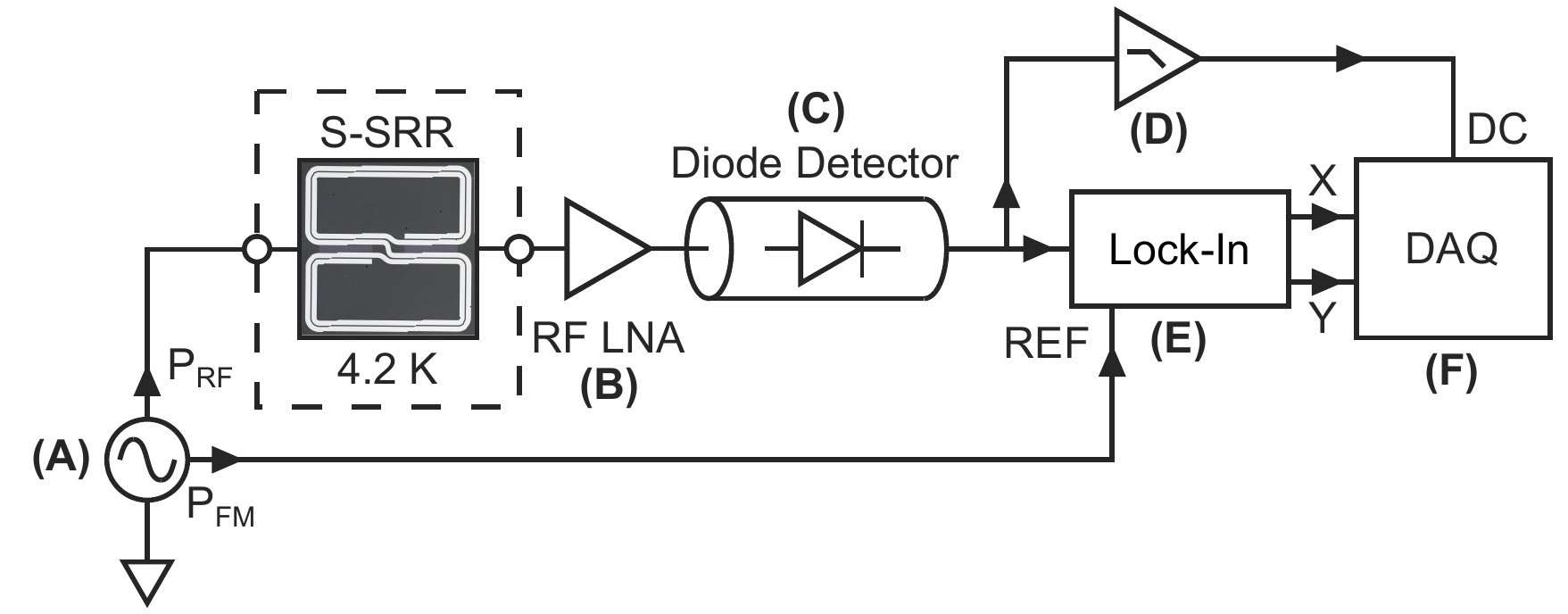}
\caption{Instrumentation setup exploited to extract the $NET$ of the different S-SRRs in liquid He at $\text{\SI{4.2}{\kelvin}}$.}
\label{ADD_INFO_04}
\end{figure*}

\autoref{ADD_INFO_04} schematically depicts the instrumentation setup. First, a microwave signal generator (\textbf{A}. \textit{Stanford Research SG384}) is exploited to excite the S-SRR with a power $P_{RF}$, frequency-modulating the excitation at $f_{FM}$ and with a $\pm \Delta f_{FM}$ deviation. The same generator additionally isolates the modulation component $P_{FM}$, for a successive synchronous lock-in detection. In liquid He, the excitation system, consisting of the Cu GCPW electromagnetically coupled to the S-SSRs, is encapsulated inside a cylindrical enclosure made of CRYOPHY\textsuperscript{\textregistered}. At room temperature, the microwave signal is amplified (\textbf{B}. \textit{Mini-Circuits ZRL-1150LN}) and converted into a DC value proportional to its amplitude by means of a microwave diode detector (\textbf{C}. \textit{Macom 2086-6000-13}). The signal is then split, with the DC component further amplified (\textbf{D}. \textit{EG\&G Princeton Applied Research Model 5113}), with a low-pass filter set to $\text{\SI{30}{\kilo\hertz}}$, and the AC one detected by means of a lock-in amplifier (\textbf{E}. \textit{EG\&G Instruments Model 7265}), with a $\text{\SI{100}{\milli\second}}$ time constant, whose reference is externally provided by the same exciting microwave generator. Finally, both DC, in-phase (X) and quadrature (Y) components are simultaneously acquired by a data acquisition (DAQ) board with a $\text{\SI{100}{\milli\second}}$ integration time (\textbf{F}. \textit{PCI-6052E}).

The following sub-sections are intended for reporting the experimental procedures adopted to identify the values of excitation power $P_{RF}$, modulation frequency $f_{FM}$ and amplitude $\Delta f_{FM}$. All the following preliminary experiments are carried out only on the $\text{\SI{147}{\nano\metre}}$ thick S-SRR.

\subsection{Excitation power $P_{RF}$ ($\Delta f_{FM}$, $f_{FM}$ fixed)}
\label{SUPPL_MAT_04_a}

The excitation microwave power $P_{RF}$ is determined along this section, by making a power sweep from $\text{\SI{-30}{\dBm}}$ to $\text{\SI{-20}{\dBm}}$ and keeping the modulation frequency and amplitude fixed (\textit{i.e.} $f_{FM} = \text{\SI{11}{\kilo\hertz}}$ and $\Delta f_{FM} = \pm\text{\SI{5}{\kilo\hertz}}$).

\begin{figure*}[h!]
\includegraphics[width=\textwidth]{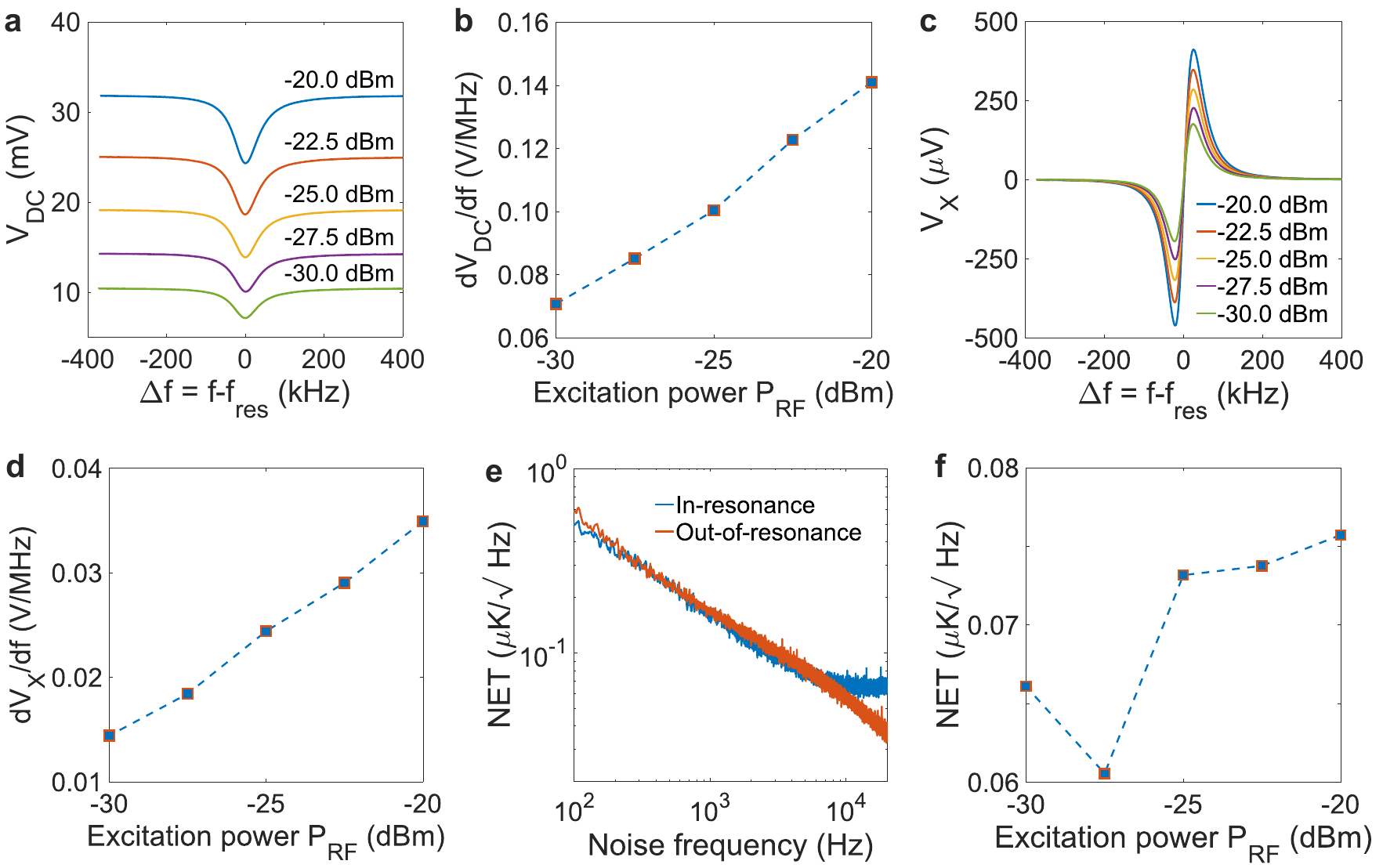}
\caption{Determination of the excitation power $P_{RF}$. a) DC resonance curve ($f_{res} = \text{\SI{1.1391}{\giga\hertz}}$) under different input powers. b) Maximum DC signal slope versus excitation power. c) X signal under different RF input powers. d) Maximum X signal slope versus excitation power. e) Frequency spectrum of the noise equivalent temperature for the DC signal at $\text{\SI{-27.5}{\dBm}}$. The blue curve represents the $NET$ extracted for the signal maximum slope condition, while the orange one stands for the out-of-resonance case. f) $NET$ evaluated at $\text{\SI{10}{\kilo\hertz}}$, under different RF excitation powers.}
\label{ADD_INFO_04_a}
\makeatletter
\let\save@currentlabel\@currentlabel
\edef\@currentlabel{\save@currentlabel(a)}\label{ADD_INFO_04_a:a}
\edef\@currentlabel{\save@currentlabel(b)}\label{ADD_INFO_04_a:b}
\edef\@currentlabel{\save@currentlabel(c)}\label{ADD_INFO_04_a:c}
\edef\@currentlabel{\save@currentlabel(d)}\label{ADD_INFO_04_a:d}
\edef\@currentlabel{\save@currentlabel(e)}\label{ADD_INFO_04_a:e}
\edef\@currentlabel{\save@currentlabel(f)}\label{ADD_INFO_04_a:f}
\edef\@currentlabel{\save@currentlabel(a-d)}\label{ADD_INFO_04_a:a-d}
\makeatother
\end{figure*}

\noindent As expected for an increase of the excitation intensity, a larger $dV/df$ curve slope is observed, increasing with the power, for both the DC and X signal components [see \autoref{ADD_INFO_04_a:a-d}], with maximum values around $\text{\SI{0.14}{\volt/\mega\hertz}}$ and $\text{ \SI{0.04}{\volt/\mega\hertz}}$ respectively. Moreover, the DC $NET$ for all the power values under test is also extracted, for the signal maximum slope case (\textit{i.e.} $\Delta f = \text{\SI{-18}{\kilo\hertz}}$), as well as for an out-of-resonance condition (\textit{i.e.} $\Delta f = \text{\SI{400}{\kilo\hertz}}$). It is interesting to notice that the coincidence of the two curves, showing a $1/f$ behavior up to a $\text{\SI{10}{\kilo\hertz}}$ corner frequency for the in-resonance signal, suggests that the noise originates from the electronic instrumentation involved [see \autoref{ADD_INFO_04_a:e}]. This result justifies the choice of FM techniques, in order to transfer low-frequency signals to frequencies above the corner, without any additional transfer of instrumentation noise. Finally, as reported in \autoref{ADD_INFO_04_a:f}, the fact that no significant difference is recorded for the minimum temperature resolutions (around $\text{\SI{0.07}{\mu\kelvin/\hertz^{1/2}}}$), in this specific power range, motivates the choice of a $\text{\SI{-20}{\dBm}}$ microwave excitation for further cryogenic tests, in order to take advantage of the higher temperature responsivity of the DC and X signals. Larger excitation powers are not considered in this work, because of possible self-heating and peak distortions due to non-linear effects.

\subsection{Modulation amplitude $\Delta f_{FM}$ ($P_{RF}$, $f_{FM}$ fixed)}
\label{SUPPL_MAT_04_b}

\begin{figure*}[t!]
\includegraphics[width=\textwidth]{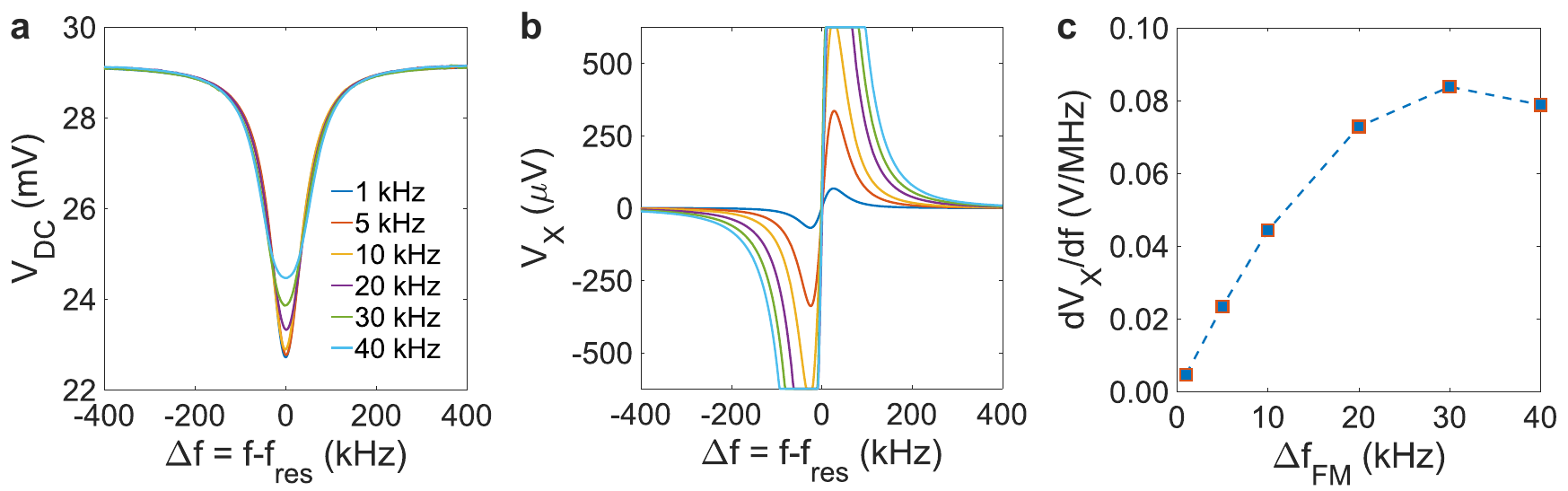}
\caption{Determination of the modulation amplitude $\Delta f_{FM}$. a) DC resonance curve ($f_{res} = \text{\SI{1.1391}{\giga\hertz}}$) under different FM amplitudes. b) X signal under different FM amplitudes: the color legend is the same as the one reported for the DC signal graph. c) Maximum X signal slope (in-resonance) versus FM amplitude.}
\label{ADD_INFO_04_b}
\makeatletter
\let\save@currentlabel\@currentlabel
\edef\@currentlabel{\save@currentlabel(a)}\label{add_info_04_b:a}
\edef\@currentlabel{\save@currentlabel(b)}\label{add_info_04_b:b}
\edef\@currentlabel{\save@currentlabel(c)}\label{add_info_04_b:c}
\edef\@currentlabel{\save@currentlabel(a-b)}\label{add_info_04_b:a-b}
\makeatother
\end{figure*}

The modulation amplitude $\Delta f_{FM}$ is determined along this section, by making a sweep from $\pm\text{\SI{5}{\kilo\hertz}}$ to $\pm\text{\SI{40}{\kilo\hertz}}$ and keeping the microwave excitation power and the modulation frequency fixed (\textit{i.e.} $P_{RF} = \text{\SI{-20}{\dBm}}$ and $f_{FM} = \text{\SI{11}{\kilo\hertz}}$). In \autoref{add_info_04_b:a-b}, both the DC and X signals versus frequency, for different FM amplitudes, are presented. In particular, the slope of the DC component is showing significant distortions, turning away from linearity, starting from $\pm\text{\SI{20}{\kilo\hertz}}$ [see \autoref{add_info_04_b:c}], due to the frequency deviation starting to be comparable with the resonance FWHM of about $\text{\SI{80}{\kilo\hertz}}$. Moreover, preliminary estimations of the $NET$, by exciting the S-SRR with higher $\Delta f_{FM}$, have not shown any reduction of the temperature resolution. As illustrated in the main text, we observe that the major sources of noise are related to the physical environment (\textit{i.e.} LHe temperature, dewar pressure, magnetic field, etc...). In such a perspective, a larger $\Delta f_{FM}$ results in the same amplification factor for both signal and environmental fluctuations, making no significant improvement in the estimation of the $NET$. An FM deviation equal to $\pm\text{\SI{5}{\kilo\hertz}}$ is, therefore, selected for performing further cryogenic tests.

\subsection{Modulation frequency $f_{FM}$ ($P_{RF}$, $\Delta f_{FM}$ fixed)}
\label{SUPPL_MAT_04_c}

\begin{figure*}[t!]
\includegraphics[width=\textwidth]{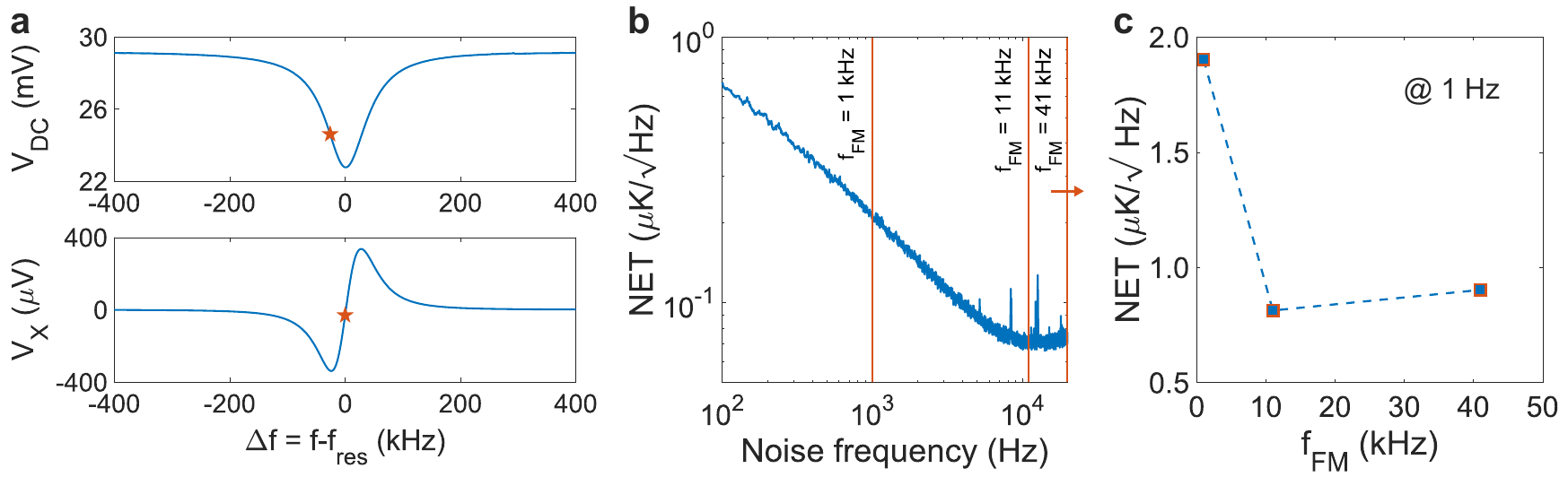}
\caption{Determination of the modulation frequency $f_{FM}$. a) DC (top) and X (bottom) resonance signals ($f_{res} = \text{\SI{1.1391}{\giga\hertz}}$). The single orange points indicate the condition of maximum slopes for both curves, where time scans get carried out to extract the respective low-frequency NSDs. b) Noise spectrum of the DC equivalent temperature, extracted while time-monitoring the signal in resonance (\textit{i.e.} for $\Delta f = \text{\SI{-18}{\kilo\hertz}}$). The orange vertical lines represent the modulation frequencies selected to perform the X-signal lock-in monitoring. c) X signal $NET$, evaluated at $\text{\SI{1}{\hertz}}$, for different choices of the modulation frequency.}
\label{ADD_INFO_04_c}
\makeatletter
\let\save@currentlabel\@currentlabel
\edef\@currentlabel{\save@currentlabel(a)}\label{ADD_INFO_04_c:a}
\edef\@currentlabel{\save@currentlabel(b)}\label{ADD_INFO_04_c:b}
\edef\@currentlabel{\save@currentlabel(c)}\label{ADD_INFO_04_c:c}
\makeatother
\end{figure*}

The modulation frequency $f_{FM}$ is determined along this section, by extracting the X-signal $NET$ for three different FMs: around $\text{\SI{1}{\kilo\hertz}}$ (well below the $1/f$-noise corner frequency), $\text{\SI{11}{\kilo\hertz}}$ (slightly above the corner frequency) and $\text{\SI{41}{\kilo\hertz}}$ (well above the corner frequency). In this case, the microwave excitation power and the modulation amplitude are kept fixed (\textit{i.e.} $P_{RF} = \text{\SI{-20}{\dBm}}$ and $\Delta f_{FM} = \pm\text{\SI{5}{\kilo\hertz}}$). \autoref{ADD_INFO_04_c:a} reports the typical response of the DC and X signals, for $f_{FM} = \text{\SI{11}{\kilo\hertz}}$, with respective maximum slopes $dV/df$ of $\text{\SI{0.10}{\volt/\mega\hertz}}$ and $\text{\SI{0.02}{\volt/\mega\hertz}}$. The DC $NET$ is presented in \autoref{ADD_INFO_04_c:b}, being acquired with a $\text{\SI{1}{\second}}$ integration time, a $\SI{200}{\kilo\hertz}$ sampling rate and a low-pass filter set to $\text{\SI{30}{\kilo\hertz}}$, to avoid possible aliasing. Since the system exhibits a $1/f$ behavior below a corner frequency around $\text{\SI{10}{\kilo\hertz}}$, an FM value situated above such a corner must be selected, in order to properly apply lock-in detection techniques to reduce the instrumentation noise. A modulation frequency equal to $\text{\SI{11}{\kilo\hertz}}$ is, therefore, selected for performing further cryogenic tests, as values below the corner condition significantly deteriorate the temperature resolution [see \autoref{ADD_INFO_04_c:c}].

\section{Low-frequency temperature fluctuations in liquid He}
\label{SUPPL_MAT_05}
In this last section, we discuss about the experiment carried out to further investigate the origin of the low-frequency temperature fluctuations. As reported in the main text, the temperature of the LHe dewar is now simultaneously monitored by exploiting two devices relying on different sensing principles: the $\text{\SI{147}{\nano\metre}}$ thick $L_K(T)$-based S-SRR and a commercial RuO\textsubscript{2} thermistor (\textit{i.e.} RO-600 by \textit{Scientific Instruments, Inc.}, showing a resistance $R(\SI{4.2}{\kelvin}) = \text{\SI{1.36}{\kilo\ohm}}$ and a temperature sensitivity $dR/dT(\SI{4.2}{\kelvin}) = \text{\SI{-86}{\ohm/\kelvin}}$, for a $\text{\SI{10}{\mu\ampere}}$ biasing current)\cite{Ihas:1998}. Considering the theoretical thermal noise limit associated to its resistance, with a $\sqrt{4k_BTR}/(I_{bias} \cdot dR/dT) \simeq \text{\SI{0.7}{\mu\kelvin/\hertz^{1/2}}}$ spectral plateau at $\text{\SI{4.2}{\kelvin}}$, such a thermistor should, in principle, allow for a sufficiently precise monitoring of the low-frequency temperature fluctuations around $\text{\SI{100}{\mu\kelvin}}$, showing a comparable resolution to the reported values for our S-SRRs.

\begin{figure*}[t!]
\includegraphics[width=\textwidth]{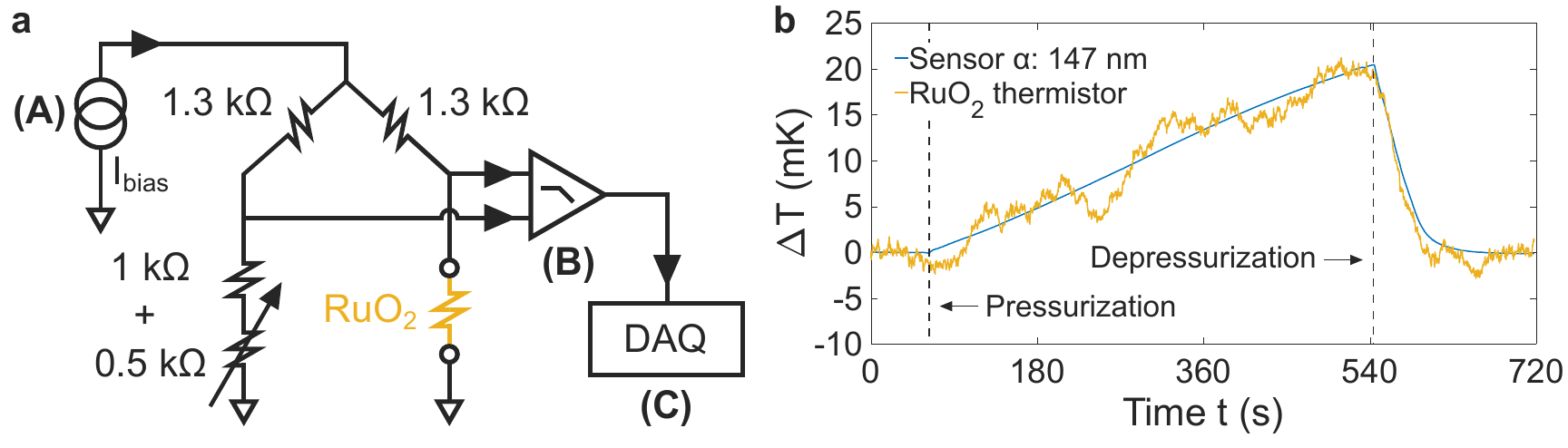}
\caption{Parallel temperature monitoring of the LHe bath at $\text{\SI{4.2}{\kelvin}}$, exploiting the $\text{\SI{147}{\nano\metre}}$ thick S-SRR and a commercial RuO\textsubscript{2} thermistor. a) DC Wheatstone bridge experimental setup used to bias the thermistor. b) Comparison of the temperature reading of the S-SRR versus the RuO\textsubscript{2} one, when performing a pressurization/depressurization of the LHe dewar.}
\label{ADD_INFO_05}
\makeatletter
\let\save@currentlabel\@currentlabel
\edef\@currentlabel{\save@currentlabel(a)}\label{ADD_INFO_05:a}
\edef\@currentlabel{\save@currentlabel(b)}\label{ADD_INFO_05:b}
\makeatother
\end{figure*}

The excitation of the S-SRR is provided by the same instrumentation setup reported in \autoref{ADD_INFO_04}, by acquiring the demodulated X-signal at the output of a lock-in amplifier. The excitation parameters are the same optimal ones identified in \autoref{SUPPL_MAT_04} (\textit{i.e.} $P_{RF} = \text{\SI{-20}{\dBm}}$, $f_{FM} = \text{\SI{11}{\kilo\hertz}}$ and $\Delta f_{FM} = \pm\text{\SI{5}{\kilo\hertz}}$).  As depicted in \autoref{ADD_INFO_05:a}, the RuO\textsubscript{2} thermistor is excited by means of a balanced Wheatstone bridge, kept at room temperature and biased through a sourcemeter (\textbf{A}. \textit{Keithley 2400}) delivering a $\text{\SI{20}{\mu\ampere}}$ DC current. A $\text{\SI{500}{\ohm}}$ potentiometer, in series to a $\text{\SI{1}{\kilo\ohm}}$ resistor, is placed on the branch opposite to the temperature-sensitive element, in order to accurately cancel out the differential voltage reading and balance the bridge. The voltage difference between the two branches is, then, amplified (\textbf{B}. \textit{EG\&G Princeton Applied Research Model 5113}), low-pass filtered, with a $\text{\SI{3}{\hertz}}$ bandwidth, and finally acquired by a DAQ board with a $\text{\SI{100}{\milli\second}}$ integration time (\textbf{C}. \textit{PCI-6052E}). The RuO\textsubscript{2} thermometer is connected, in a 4-wires configuration, to the DC pads of the Cu PCB and placed at a distance of about $\text{\SI{1}{\centi\metre}}$ far from the S-SRR.

A preliminary experiment is carried out in order to check the correct functioning of the sensors. A relatively large pressure-induced temperature perturbation of about $\text{\SI{20}{\milli\kelvin}}$ is generated by closing and opening the valve connecting the LHe dewar to a gas recovery line. In \autoref{ADD_INFO_05:b} we present the monitoring of the temperature along this test. The temperature of the LHe bath increases linearly upon closing the valve, because of the increase of the internal vapor pressure. When opening the valve, the systems relaxes back to the initial situation through an exponential decay. Moreover, we also record an almost $\text{\SI{20}{\milli\bar}}$ increase in the internal pressure of the dewar, by means of a barometer placed on the head of the dewar, which should ideally correspond to a temperature increase of about $\text{\SI{20}{\milli\kelvin}}$ for boiling liquid He at $\text{\SI{4.2}{\kelvin}}$\cite{Donnelly:1998}. The coincidence of the two responses, upon voltage-to-temperature conversion, as well as the value of the overall temperature shift, is an indicator of the reliability of the temperature reading through both sensors. The noise level of the thermistor (about $\text{\SI{2}{\milli\kelvin}}$) turns out to be almost 400 times higher than the one associated to the resonator (about $\text{\SI{5}{\mu\kelvin}}$). Such a relatively poor temperature resolution, for the DC-biased thermistor, would theoretically prevent the ultimate detection of the low-frequency temperature fluctuations, in the order of  $\text{\SI{100}{\mu\kelvin}}$.\\
A lock-in detection scheme is, therefore, implemented in an attempt to further reduce the noise associated to the electronics involved to monitor the LHe bath temperature by means of the RuO\textsubscript{2} thermistor. Such a final experiment is fully described in the main text.

\printbibliography[heading=bibintoc]